\documentclass[times,preprint,3p]{elsarticle}

\usepackage{framed,multirow}
\usepackage{float}

\usepackage{bm}
\usepackage{multirow}
\usepackage{hyperref}
\usepackage{color}
\usepackage{amsfonts}
\usepackage{amsmath}
\usepackage{float}
\usepackage{graphicx}
\usepackage{rotating}
\usepackage{mathtools}
\usepackage[shortlabels]{enumitem}
\usepackage{siunitx} 
\usepackage{graphicx}
\usepackage{subcaption}
\usepackage{hhline}
\usepackage{hyperref}
\usepackage[capitalise]{cleveref}

\hypersetup{
    colorlinks=true,
    linkcolor=blue,
    filecolor=magenta,      
    urlcolor=cyan,
}

\usepackage{booktabs}
\usepackage{lineno}

\makeatletter
\def\@xfootnote[#1]{%
  \protected@xdef\@thefnmark{#1}%
  \@footnotemark\@footnotetext}
\makeatother

\usepackage{url}
\usepackage{xcolor}
\definecolor{newcolor}{rgb}{.8,.349,.1}

\newcommand{\atcell}[2]{
  \left.#1\right|_{#2}
}
\newcommand{\atcelltime}[3]{
  \left.#1\right|_{#2}^{#3}
}
\newcommand{\atfaceplus}[2]{
  \left.#1\right|_{#2 + \frac{1}{2}}
}
\newcommand{\atfaceminus}[2]{
  \left.#1\right|_{#2 - \frac{1}{2}}
}
\newcommand{\atfaceplustime}[3]{
  \left.#1\right|_{#2 + \frac{1}{2}}^{#3}
}
\newcommand{\atfaceminustime}[3]{
  \left.#1\right|_{#2 - \frac{1}{2}}^{#3}
}

\begin{document}
\begin{frontmatter}

\title{On the Construction of High-Order and Exact Pressure Equilibrium Schemes for Arbitrary Equations of State}

\author[1,3]{Christopher DeGrendele}
\ead{christopher.j.degrendele@nasa.gov}

\author[2]{Nguyen Ly}
\ead{nguyen.ly@nasa.gov}

\author[1]{Francois Cadieux}

\author[1]{Michael Barad}

\author[3]{Dongwook {Lee}}

\author[1]{Jared Duensing}

\address[1]{NASA Ames Research Center, Moffett Field, CA, United States}
\address[2]{Analytical Mechanics Associates, Moffett Field, CA, United States}
\address[3]{Department of Applied Mathematics, 
The University of California, Santa Cruz, CA, United States}

\begin{abstract}
Typical fully conservative discretizations of the Euler compressible single or multi-component fluid equations governed by a real-fluid equation of state exhibit spurious pressure oscillations due to the nonlinearity of the thermodynamic relation between pressure, density, and internal energy. A fully conservative, pressure-equilibrium preserving method and a high-order, fully conservative, approximate pressure-equilibrium preserving method are presented. Both methods are general and can handle an arbitrary equation of state and arbitrary number of species. Unlike existing approaches to discretize the multi-component Euler equations, we do not introduce non conservative updates, overspecified equations, or design for a specific equation of state. The proposed methods are demonstrated on inviscid smooth interface advection problems governed by three equations of state: ideal-gas, stiffened-gas, and van der Waals where we show orders of magnitude reductions in spurious pressure oscillations compared to existing schemes. 
\end{abstract}

\end{frontmatter}


\section{Introduction}
Spurious pressure oscillations have been observed and studied in multi-component and real-fluid Euler simulations for decades \cite{abgrall_computations_2001}. The root cause of this error is well known: when the ratio of specific heats is not constant across species, or when, in real fluids, the effective \(\gamma\) varies with local thermodynamic conditions, small inconsistencies enter the discrete thermodynamic state at the cell interface. In practice the resulting pressure error is of the order of the discretization error per step, but it can and will grow indefinitely \cite{fujiwara_fully_2023}. For simulations with long time horizons or with uniform-pressure initial data, this growth will eventually dominate the solution, making it a problem that must be addressed. In this paper we focus on the inviscid, multi-component compressible Euler equations. Because the mechanism is a discrete thermodynamic inconsistency at material interfaces when effective thermodynamic parameters vary, spurious pressure oscillations will also arise in other four-equation mixture models and in viscous extensions, although physical/numerical diffusion may partially damp their growth.

A number of approaches have been developed to preserve the pressure-equilibrium property (PEP). A simple option is to adopt a non-conservative update in which the numerical flux update is applied directly to the pressure field \cite{karni_hybrid_1996}. This enforces pressure equilibrium exactly at the discrete level, but it does so at the cost of strict conservation of total energy. More recently, \cite{ching2025conservativepressureequilibriumpreservingdiscontinuousgalerkin} proposed a discontinuous Galerkin scheme that likewise evolves a pressure-evolution equation in place of the total-energy equation, and then adds Abgrall-type correction terms so that the final method is velocity- and pressure-equilibrium-preserving and (semi-discretely) energy-conservative for mixtures of thermally perfect gases. A related and widely used idea, often referred to as the double-flux method \cite{ma_entropy-stable_2017}, updates the conservative variables and, in addition, evolves a pressure flux that is used as a correction to the total energy. In practice this aligns the energy update more closely with the pressure evolution and generally reduces energy drift, though the scheme remains formally non-conservative. There are also quasi-conservative formulations \cite{abgrall_how_1996, liu_quasi-conservative_1998} in which the governing equations are modified to evolve an auxiliary term related to the effective ratio of specific heats; for ideal gases this can restore PEP under the altered dynamics. Some systems, such as the five-equation model for multiphase flows \cite{allaire_five-equation_2002}, satisfy PEP by construction and require no additional treatment. Complementary mechanisms include consistent artificial dissipation designed to damp spurious pressure modes without corrupting equilibrium states \cite{terashima_consistent_2013}, and hybrid schemes that switch between quasi-conservative and double-flux updates to balance robustness and conservation locally \cite{boyd_diffuse-interface_2021}.

More recently, \cite{fujiwara_fully_2023} presented the first fully conservative scheme that numerically satisfies PEP for the ideal-gas and stiffened-gas equations of state (with zero reference energy \(q\)). Their analysis identifies an exact algebraic relation coupling the discrete species and energy updates so that a uniform-pressure state is an invariant of the semi-discrete system during contact advection. Building on the same principle, \cite{TERASHIMA2025113701} derived an approximate pressure-equilibrium conservative (APEC) scheme for general equations of state (EOS) by enforcing an approximate form of that algebraic condition. As formulated, APEC substantially reduces pressure oscillations for complex EOS, but the original presentation is limited to a second-order scheme in space.

In this work we take two steps. First, we extend APEC to high-order by formulating interface-consistent energy corrections that are compatible with high-order reconstruction and flux-differencing, removing the second-order limitation while retaining the intended pressure stabilization. Second, we derive, for an arbitrary EOS, the first fully conservative scheme that exactly \emph{spatially} satisfies the PEP condition at the semi-discrete level. The resulting discretization preserves conservation of mass, momentum, and energy while significantly minimizing accrued pressure error over time.

This paper is organized as follows. Section~\ref{sec:derivation_pe} derives the discrete pressure–equilibrium condition, Sections~\ref{sec:high_order_apec} and \ref{sec:fully_conservative_PEP} develop high-order APEC and a fully conservative PEP scheme for general EOS, and Section~\ref{sec:results} presents numerical results for ideal-gas, stiffened-gas, and van der Waals mixtures.

\section{Derivation of the Pressure Equilibrium Preserving Condition}
\label{sec:derivation_pe}
For the purpose of the derivation, we consider the 1D Euler equation for two species written as
\begin{equation}
\begin{aligned}
    \frac{\partial (\rho Y_1)}{\partial t} + \frac{\partial (\rho Y_1 u)}{\partial x} &= 0 \\
    \frac{\partial (\rho Y_2)}{\partial t} + \frac{\partial (\rho Y_2 u)}{\partial x} &= 0 \\
    \frac{\partial (\rho u)}{\partial t} + \frac{\partial (\rho u u + p)}{\partial x} &= 0 \\
    \frac{\partial (\rho E)}{\partial t} + \frac{\partial \left[(\rho E + p) u\right]}{\partial x} &= 0,
\end{aligned}
\end{equation}
where $E = e + u^2/2$

We will base our derivation of the discrete spatial pressure-equilibrium criterion on the approach of Fujiwara \cite{fujiwara_fully_2023}, with some alterations to explicitly demonstrate the temporal convergence aspect of the pressure-equilibrium scheme. Consider a solution of the multispecies 1D Euler equation using the method of lines, where a spatial discretization is employed to convert the PDE system into an ODE system of the discretized flow variables in time
\begin{gather}
    \atcell{\frac{d (\rho Y_1)}{d t}}{m} + \atcell{F_{\rho Y_1}}{m} = 0 \\
    \atcell{\frac{d (\rho Y_2)}{d t}}{m} + \atcell{F_{\rho Y_2}}{m} = 0 \\
    \atcell{\frac{d (\rho u)}{d t}}{m} + \atcell{F_{\rho u}}{m} = 0 \\
    \atcell{\frac{\partial (\rho E)}{\partial t}}{m} +  \atcell{F_{\rho E}}{m} = 0
\end{gather}
where $\atcell{F_{\rho Y_1}}{m}, \atcell{F_{\rho Y_2}}{m}, \atcell{F_{\rho u}}{m}, \atcell{F_{\rho E}}{m}$ are the discretized convective flux operators and $m$ is the spatial grid point index. In this study, we consider the split form of the Kinetic Energy Entropy Preserving (KEEP) scheme \cite{kuya_kinetic_2018} for these convective flux operators
\begin{align}
    \atcell{F_{\rho Y_1}}{m} &= \frac{\atfaceplus{C_1}{m} - \atfaceminus{C_1}{m}}{\Delta x} \\
    \atcell{F_{\rho Y_2}}{m} &= \frac{\atfaceplus{C_2}{m} - \atfaceminus{C_2}{m}}{\Delta x} \\
    \atcell{F_{\rho u}}{m} &= \frac{\atfaceplus{M}{m} - \atfaceminus{M}{m}}{\Delta x} +  \frac{\atfaceplus{\Pi}{m} - \atfaceminus{\Pi}{m}}{\Delta x} \\
    \atcell{F_{\rho E}}{m} &=  \frac{\atfaceplus{K}{m} - \atfaceminus{K}{m}}{\Delta x} + \frac{\atfaceplus{I}{m} - \atfaceminus{I}{m}}{\Delta x} + \frac{\atfaceplus{S}{m} - \atfaceminus{S}{m}}{\Delta x},
\end{align}
where
\begin{align}
    \atfaceplus{C_1}{m} &= \atfaceplus{\rho Y_1}{m} \frac{\atcell{u}{m} + \atcell{u}{m+1}}{2} \label{eq:keep_flux_beg}\\
    \atfaceplus{C_2}{m} &= \atfaceplus{\rho Y_2}{m} \frac{\atcell{u}{m} + \atcell{u}{m+1}}{2} \\
    \atfaceplus{M}{m} &= \atfaceplus{\rho}{m} \frac{\atcell{u}{m} + \atcell{u}{m+1}}{2} \frac{\atcell{u}{m} + \atcell{u}{m+1}}{2} \\
    \atfaceplus{\Pi}{m} &= \frac{\atcell{p}{m} + \atcell{p}{m+1}}{2} \\
    \atfaceplus{K}{m} &= \atfaceplus{\rho}{m} \frac{\atcell{u}{m} + \atcell{u}{m+1}}{2} \frac{\atcell{u}{m} \atcell{u}{m+1}}{2} \\
    \atfaceplus{I}{m} &= \atfaceplus{\rho e}{m} \frac{\atcell{u}{m} + \atcell{u}{m+1}}{2} \\
    \atfaceplus{S}{m} &= \frac{\atcell{u}{m} \atcell{p}{m+1} + \atcell{u}{m+1} \atcell{p}{m}}{2}
    \label{eq:keep_flux_end}
\end{align}

We refer to this baseline numerical method as our Non-Pressure-Equilbrium method ``NPE". In the method of line, the conserved variables $\rho Y_1, \rho Y_2, \rho u, \rho E$ are directly updated using time-stepping of the semi-discrete ODEs. On the other hand, the derived variable $p$ is indirectly updated from the conserved variables via
\begin{equation}
    p = \text{EOS}(\rho e,\rho Y_1, \rho Y_2) = \text{EOS}\left(\rho E - \frac{1}{2}\frac{(\rho u)^2}{\rho}, \rho Y_1, \rho Y_2\right)
\end{equation}

The pressure-equilibrium preserving property is analyzed for an initial flow state with constant velocity and pressure: $u(x,0) = U_0$ and $p(x,0) = p_0$. First, we analyze the velocity-equilibrium aspect of the scheme. Analyzing velocity equilibrium first ensures that no artificial momentum errors propagate into the pressure field, allowing a clean assessment of the pressure-equilibrium condition. Assuming a single-step temporal scheme for the update of flow variables from time $t_n$ to time $t_{n+1}$, then
\begin{gather}
    \atcelltime{u}{m}{n+1} - \atcelltime{u}{m}{n} = \atcelltime{\left(\frac{\partial u}{\partial \rho u}\right)_\rho}{m}{n} (\atcelltime{\rho u}{m}{n+1} - \atcelltime{\rho u}{m}{n}) + \atcelltime{\left(\frac{\partial u}{\partial \rho}\right)_{\rho u}}{m}{n} (\atcelltime{\rho}{m}{n+1} - \atcelltime{\rho}{m}{n}) + \text{H.O.} \\
    = -\frac{1}{\atcelltime{\rho}{m}{n}} \atcelltime{F_{\rho u}}{m}{n} \Delta t + \frac{U_0}{\atcelltime{\rho}{m}{n}} (\atcelltime{F_{\rho Y_1}}{m}{n} + \atcelltime{F_{\rho Y_2}}{m}{n}) \Delta t + O(\Delta t^2).
\end{gather}
Here, we only collect the leading order terms in the time stepping, leading to a first-order global convergence of the velocity equilibrium property. The same treatment will be made for the pressure equilibrium property later, with further remarks. Substituting the KEEP scheme into the (implicit) velocity update rule, under the assumption of constant velocity and pressure at time step $t_n$, gives
\begin{gather}
    \atcelltime{u}{m}{n+1} - \atcelltime{u}{m}{n} = \frac{U_0^2}{\atcell{\rho}{m}}\left[-\left(\atfaceplustime{\rho}{m}{n} - \atfaceminustime{\rho}{m}{n}\right) + \left(\atfaceplustime{\rho Y_1}{m}{n} - \atfaceminustime{\rho Y_1}{m}{n}\right) + \left(\atfaceplustime{\rho Y_2}{m}{n} - \atfaceminustime{\rho Y_2}{m}{n}\right)\right]\frac{\Delta t}{\Delta x} + O(\Delta t^2).
\end{gather}
From here, we see that with the choice of mid-point density
\begin{equation}
    \atfaceplus{\rho}{m} \equiv \atfaceplus{\rho Y_1}{m} +  \atfaceplus{\rho Y_2}{m},
\end{equation}
the velocity equilibrium is preserved with first order global convergence in time \emph{for all grid spacing}. In this manner, we refer to the present scheme as being \emph{spatially exact} in preserving velocity equilibrium. Further remarks on this point will be made shortly when the same treatment is considered for the pressure.

For pressure, since the update will consider the internal energy through the EOS, we first look at the implicit update rule for $\rho e = \rho E - \frac{1}{2}(\rho u)^2 / \rho$, once again assuming a single-step temporal scheme,

\begin{equation}
    \begin{aligned}
        \atcelltime{\rho e}{m}{n+1} - \atcelltime{\rho e}{m}{n}
        &= \atcelltime{\left(\frac{\partial \rho e}{\partial \rho E}\right)_{\rho u, \rho}}{m}{n} \big(\atcelltime{\rho E}{m}{n+1} - \atcelltime{\rho E}{m}{n}\big) 
         +\,\atcelltime{\left(\frac{\partial \rho e}{\partial (\rho u)}\right)_{\rho E, \rho}}{m}{n} \big(\atcelltime{\rho u}{m}{n+1} - \atcelltime{\rho u}{m}{n}\big) \\
        & +\,\atcelltime{\left(\frac{\partial \rho e}{\partial \rho}\right)_{\rho E, \rho u}}{m}{n} \big(\atcelltime{\rho}{m}{n+1} - \atcelltime{\rho}{m}{n}\big)
        + \text{H.O.}
    \end{aligned}
\end{equation}

\begin{equation}
    = -\,\atcelltime{F_{\rho E}}{m}{n}\,\Delta t
+\,\atcelltime{u}{m}{n}\,\atcelltime{F_{\rho u}}{m}{n}\,\Delta t
-\,\frac{1}{2}\,(\atcelltime{u}{m}{n})^2\,\atcelltime{F_\rho}{m}{n}\,\Delta t
+ O(\Delta t^2)
\label{eq:time_inexact_pressure}
\end{equation}

From Equation \ref{eq:time_inexact_pressure}, we now substitute the split-form KEEP fluxes in Equations \ref{eq:keep_flux_beg}--\ref{eq:keep_flux_end}. We then regroup the resulting face contributions into an internal-energy advection part and two ``cancellation'' parts, which vanish identically under a uniform-velocity and uniform-pressure state at $t_n$.

\begin{equation}
\label{eq:26_to_27_bridge}
\begin{aligned}
\atcelltime{\rho e}{m}{n+1} - \atcelltime{\rho e}{m}{n}
&=
-\Bigg[
\atcelltime{F_{\rho E}}{m}{n}
-\atcelltime{u}{m}{n}\,\atcelltime{F_{\rho u}}{m}{n}
+\frac{1}{2}\big(\atcelltime{u}{m}{n}\big)^2\,\atcelltime{F_{\rho}}{m}{n}
\Bigg]\Delta t
+O(\Delta t^2) \\
&=
-\frac{\Delta t}{\Delta x}\Bigg[
\Big(\atfaceplustime{I}{m}{n}-\atfaceminustime{I}{m}{n}\Big) \\
&\qquad
+\Big(\atfaceplustime{S}{m}{n}-\atfaceminustime{S}{m}{n}\Big)
-\atcelltime{u}{m}{n}\Big(\atfaceplustime{\Pi}{m}{n}-\atfaceminustime{\Pi}{m}{n}\Big) \\
&\qquad
+\Big(\atfaceplustime{K}{m}{n}-\atfaceminustime{K}{m}{n}\Big)
-\atcelltime{u}{m}{n}\Big(\atfaceplustime{M}{m}{n}-\atfaceminustime{M}{m}{n}\Big) \\
&\qquad
+\frac{1}{2}\big(\atcelltime{u}{m}{n}\big)^2
\Big(\atfaceplustime{(C_1+C_2)}{m}{n}-\atfaceminustime{(C_1+C_2)}{m}{n}\Big)
\Bigg]
+O(\Delta t^2).
\end{aligned}
\end{equation}

Assuming a constant velocity, $U_0$ and pressure $p_0$ at $t_n$ in Equation \ref{eq:26_to_27_bridge}, we get
\begin{equation}
    \atcelltime{\rho e}{m}{n+1} - \atcelltime{\rho e}{m}{n} = -U_0 \left(\atfaceplustime{\rho e}{m}{n} - \atfaceminustime{\rho e}{m}{n}\right)\frac{\Delta t}{\Delta x} + O(\Delta t^2).
\end{equation}

Now, the implicit update rule for pressure can be derived for a single-step temporal scheme,
\begin{equation}
\begin{aligned}
    \atcelltime{p}{m}{n+1} - \atcelltime{p}{m}{n} &= \atcelltime{\left(\frac{\partial p}{\partial \rho e}\right)_{\rho Y_1, \rho Y_2}}{m}{n} \left(\atcelltime{\rho e}{m}{n+1} - \atcelltime{\rho e}{m}{n}\right) + \atcelltime{\left(\frac{\partial p}{\partial \rho Y_1}\right)_{\rho e, \rho Y_2}}{m}{n} \left(\atcelltime{\rho Y_1}{m}{n+1} - \atcelltime{\rho Y_1}{m}{n}\right)  \\
    &+ \atcelltime{\left(\frac{\partial p}{\partial \rho Y_2}\right)_{\rho e, \rho Y_1}}{m}{n} \left(\atcelltime{\rho Y_2}{m}{n+1} - \atcelltime{\rho Y_2}{m}{n}\right) + \text{H.O.} 
\end{aligned}
\end{equation}

\begin{equation}
\begin{aligned}
    &= -U_0 \left[\atcelltime{\left(\frac{\partial p}{\partial \rho e}\right)_{\rho Y_1, \rho Y_2}}{m}{n} \left(\atfaceplustime{\rho e}{m}{n} - \atfaceminustime{\rho e}{m}{n}\right) +  \atcelltime{\left(\frac{\partial p}{\partial \rho Y_1}\right)_{\rho e, \rho Y_2}}{m}{n} \left(\atfaceplustime{\rho Y_1}{m}{n} - \atfaceminustime{\rho Y_1}{m}{n}\right) \right. \\
    &+  \left. \atcelltime{\left(\frac{\partial p}{\partial \rho Y_2}\right)_{\rho e, \rho Y_1}}{m}{n} \left(\atfaceplustime{\rho Y_2}{m}{n} - \atfaceminustime{\rho Y_2}{m}{n}\right)\right] \frac{\Delta t}{\Delta x} + O(\Delta t^2).
\end{aligned}
\end{equation}

Thus, we have derived the condition for spatially exact pressure equilibrium preserving. This criterion is expressed in terms of the mid-point values for internal energy and the partial densities and is identical in form to the criterion derived in \cite{fujiwara_fully_2023},  with the added context of temporal convergence,
\begin{equation}
\begin{aligned}
    &\atcelltime{\left(\frac{\partial p}{\partial \rho e}\right)_{\rho Y_1, \rho Y_2}}{m}{n} \left(\atfaceplustime{\rho e}{m}{n} - \atfaceminustime{\rho e}{m}{n}\right) +  \atcelltime{\left(\frac{\partial p}{\partial \rho Y_1}\right)_{\rho e, \rho Y_2}}{m}{n} \left(\atfaceplustime{\rho Y_1}{m}{n} - \atfaceminustime{\rho Y_1}{m}{n}\right)  \\
    &+  \atcelltime{\left(\frac{\partial p}{\partial \rho Y_2}\right)_{\rho e, \rho Y_1}}{m}{n} \left(\atfaceplustime{\rho Y_2}{m}{n} - \atfaceminustime{\rho Y_2}{m}{n}\right) = 0 
\end{aligned}
\label{eq:pep_property}
\end{equation}

To rewrite Equation \ref{eq:pep_property} as a condition on the mid-point jump in $\rho e$, we solve for $\left(\atfaceplustime{\rho e}{m}{n} - \atfaceminustime{\rho e}{m}{n}\right)$ and express the resulting coefficients using the EOS along with a constant pressure. This yields
\begin{equation}
\label{eq:pep_property_bridge}
\left(\atfaceplustime{\rho e}{m}{n} - \atfaceminustime{\rho e}{m}{n}\right)
=
-\sum_{i=1}^{N}
\left.
\frac{\left(\frac{\partial p}{\partial \rho Y_i}\right)_{\rho e,\,\rho Y_{j\neq i}}}
     {\left(\frac{\partial p}{\partial \rho e}\right)_{\rho Y_1,\dots,\rho Y_N}}
\right|_{m}^{n}
\left(\atfaceplustime{\rho Y_i}{m}{n} - \atfaceminustime{\rho Y_i}{m}{n}\right)
=
\sum_{i=1}^{N}
\left.\left(\frac{\partial \rho e}{\partial \rho Y_i}\right)_{\rho Y_{j\neq i},\,p}\right|_{m}^{n}
\left(\atfaceplustime{\rho Y_i}{m}{n} - \atfaceminustime{\rho Y_i}{m}{n}\right).
\end{equation}

Finally, we write the PEP criterion as

\begin{equation}
\label{eq:pep_property_final}
   \Rightarrow \left.\rho e\right|_{m+\frac{1}{2}}-\left.\rho e\right|_{m-\frac{1}{2}}
    =\left.\sum_{i=1}^N\left(\frac{\partial \rho e}{\partial \rho Y_i}\right)_{\rho Y_{j \neq i}, p}\right|_m
\left(\left.\rho Y_i\right|_{m+\frac{1}{2}}-\left.\rho Y_i\right|_{m-\frac{1}{2}}\right).
\end{equation}
We emphasize that the pressure-equilibrium criterion derived above is only \emph{spatially} exact, in the sense that pressure error converges with the same (linear) rate with time step size for all mesh spacings. This is a common property of pressure-equilibrium preserving conservative methods \cite{fujiwara_fully_2023} since we only explicitly update the conservative variables $(\rho Y_i, \rho u, \rho E)$. Nonetheless, relative to the NPE discretization, the present approach removes the need to rely on mesh refinement to recover pressure equilibrium, since the spatial PEP condition is enforced independently of $\Delta x$. Any remaining pressure error in the fully discrete scheme is then attributable to the time discretization and is proportional to the time-step size, i.e., $O(\Delta t)$ for the single-step update used in this derivation. Higher-order temporal variants can be constructed by applying standard multi-stage or multi-step integrators to the same Euler-type updates presented here, which would yield $O(\Delta t^q)$ temporal error for a $q$th-order method. Extensions of the above analysis to higher-order truncations in $\Delta t$ will be the subject of future work.

To apply Equation \ref{eq:pep_property_final} we first split into a left and right condition shown as

\begin{equation}
\left.\rho e\right|_{m}-\left.\rho e\right|_{m-\frac{1}{2}}
=\left.\sum_{i=1}^N\left(\frac{\partial \rho e}{\partial \rho Y_i}\right)_{\rho Y_{j \neq i}, p}\right|_{m}
\left(\left.\rho Y_i\right|_{m}-\left.\rho Y_i\right|_{m-\frac{1}{2}}\right)
\label{eq:exact_condition_left}
\end{equation}

\begin{equation}
\left.\rho e\right|_{m+\frac{1}{2}}-\left.\rho e\right|_{m}
=\left.\sum_{i=1}^N\left(\frac{\partial \rho e}{\partial \rho Y_i}\right)_{\rho Y_{j \neq i}, p}\right|_{m}
\left(\left.\rho Y_i\right|_{m+\frac{1}{2}}-\left.\rho Y_i\right|_{m}\right)
\label{eq:exact_condition_right}
\end{equation}

We then shift Equation \ref{eq:exact_condition_left} by one cell to the right to get

\begin{equation}
\left.\rho e\right|_{m+1}-\left.\rho e\right|_{m+\frac{1}{2}}
=\left.\sum_{i=1}^N\left(\frac{\partial \rho e}{\partial \rho Y_i}\right)_{\rho Y_{j \neq i}, p}\right|_{m+1}
\left(\left.\rho Y_i\right|_{m+1}-\left.\rho Y_i\right|_{m+\frac{1}{2}}\right).
\label{eq:exact_condition_left_shifted}
\end{equation}

where in practice for the development of our scheme, Equations \ref{eq:exact_condition_right} and \ref{eq:exact_condition_left_shifted} represent the left and right interface conditions one must satisfy to preserve the pressure-equilibrium property.

\section{Derivation of High-Order APEC}
\label{sec:high_order_apec}

The Approximately Pressure-Equilibrium-preserving Conservative scheme (APEC) \cite{TERASHIMA2025113701} is derived by considering an approximation to the PEP criterion (Equation \ref{eq:pep_property_final}) rather than satisfying Equations \ref{eq:exact_condition_right} and \ref{eq:exact_condition_left_shifted} directly. This is done by subtracting Equation \ref{eq:exact_condition_right} from Equation \ref{eq:exact_condition_left_shifted} and solving for the half point value $\atfaceplus{\rho e}{m}$ which gives us:

\begin{equation}
\label{eq:apec_rhoe_flux}
\atfaceplus{\rho e}{m}^{\text{APEC}} = \frac{\left.\rho e\right|_{m}+\left.\rho e\right|_{m+1}}{2}
+\frac{1}{2}\left.\sum_{i=1}^N\left[\left(\frac{\partial \rho e}{\partial \rho Y_i}\right)_{\rho Y_{j \neq i}, p}\right|_{m}
\left(\left.\rho Y_i\right|_{m + \frac{1}{2}}-\left.\rho Y_i\right|_{m}\right) -
\left.\left(\frac{\partial \rho e}{\partial \rho Y_i}\right)_{\rho Y_{j \neq i}, p}\right|_{m+1}
\left(\left.\rho Y_i\right|_{m + 1}-\left.\rho Y_i\right|_{m + \frac{1}{2}}\right)\right],
\end{equation}

where $N$ is the number of species. This condition is a necessary but not sufficient condition to satisfy the PEP property and thus it is an approximate method. Terashima \cite{TERASHIMA2025113701} couples this newly derived halfpoint value with the KEEP scheme and shows a considerable decrease in pressure error and an increase in stability for an arbitrary EOS. Importantly, this correction term can also be applied to any numerical flux function, but the scheme is left at second order. We will now present the derivation of a high-order variant of this scheme.

We begin by presenting a high-order extension of the KEEP fluxes from \cite{kuya_high-order_2021}, expressed in split two-point flux form. This approach follows the general methodology first introduced in \cite{pirozzoli_generalized_2010} for extending any split convective operator to higher orders. For a general scalar variable $\phi$ in a convective term $\rho\phi u$ the numerical flux to an arbitrary order of accuracy is defined as 

\begin{equation}
\label{eq:high_order_flux}    
\left.\widetilde{\mathcal{F}}_\phi\right|_{\left(m \pm \frac{1}{2}\right)} \equiv 2 \sum_{i_s=1}^{q / 2} a_{q, i_s} \sum_{\ell=0}^{i_s-1} \frac{ \left.\rho\right|_{(m \mp \ell)}+\left.\rho \right|_{\left(m \mp \ell \pm i_s\right)}}{2} \frac{\left. \phi \right|_{(m \mp \ell)}+ \left. \phi \right|_{\left(m \mp \ell \pm i_s\right)}}{2} \frac{\left. u \right| _{(m \mp \ell)}+ \left. u \right|_{\left(m \mp \ell \pm i_s\right)}}{2},
\end{equation}

where $q$ is the order of the scheme, $i_s$ is the size of the sub-stencil, and $a_{q, i_s}$ are the central difference approximation coefficients as shown in Table \ref{table:coeffs}, 

\begin{table}[h]
\centering
\renewcommand{\arraystretch}{1.8}
\begin{tabular}{c|cccc}
\toprule
Order & \multicolumn{4}{c}{Coefficients} \\
$q$ & $a_{q,1}$ & $a_{q,2}$ & $a_{q,3}$ & $a_{q,4}$ \\
\midrule
2 & $\dfrac{1}{2}$ & & & \\[3pt]
4 & $\dfrac{2}{3}$ & $-\dfrac{1}{12}$ & & \\[3pt]
6 & $\dfrac{3}{4}$ & $-\dfrac{3}{20}$ & $\dfrac{1}{60}$ & \\[3pt]
8 & $\dfrac{4}{5}$ & $-\dfrac{1}{5}$ & $\dfrac{4}{105}$ & $-\dfrac{1}{280}$ \\
\bottomrule
\end{tabular}
\caption{Coefficients ($a_{q,i_s}$) for central difference approximation of first derivative as shown in \cite{kuya_high-order_2021} and derived in \cite{fornberg_generation_1988}. $q$ is the order of accuracy of the method and $i_s$ is the stencil size.}
\label{table:coeffs}
\end{table}

\begin{figure}
    \centering
    \includegraphics[width=\linewidth]{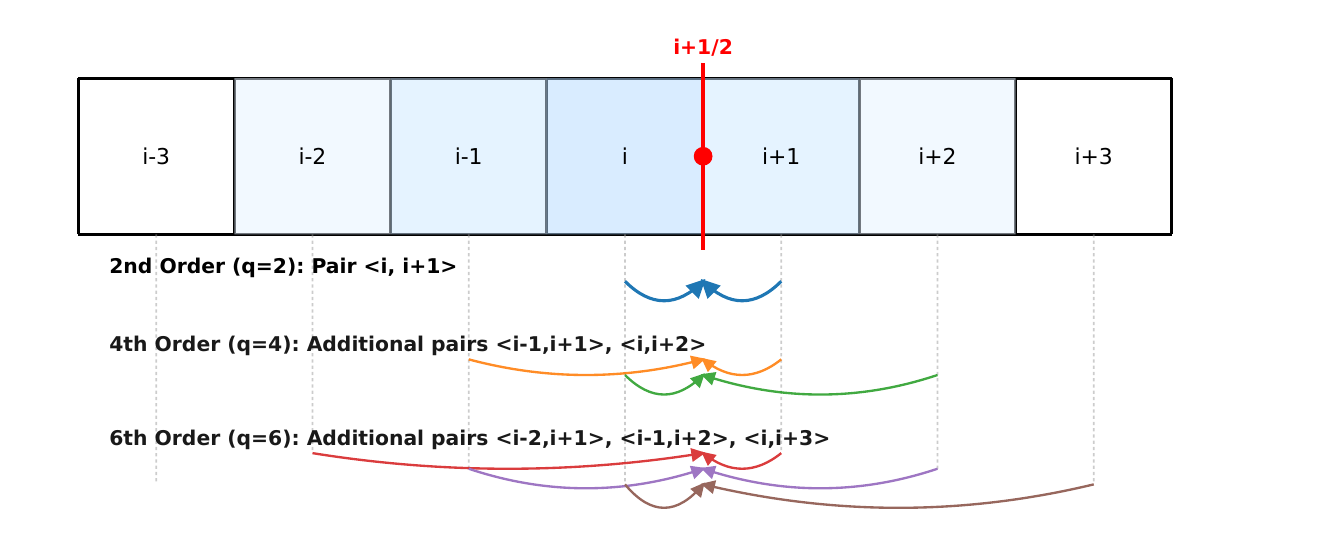}
    \caption{Construction of the high-order interface flux at $i+1/2$ as a weighted sum of multiple cell-pair contributions. Dashed gray lines connect cell centers to their arrow origins. For second-order accuracy (blue), a single pair $\langle i, i+1 \rangle$ forms the sub-interface state. Fourth-order accuracy adds two additional pairs: $\langle i-1, i+1 \rangle$ (orange) and $\langle i, i+2 \rangle$ (green). Sixth-order accuracy further extends the stencil with three more pairs: $\langle i-2, i+1 \rangle$ (red), $\langle i-1, i+2 \rangle$ (purple), and $\langle i, i+3 \rangle$ (brown). Each arrow pair represents a sub-interface contribution formed by averaging the pointwise values at two cell centers, and the final interface flux is the weighted sum of all these sub-interface contributions according to Equation \ref{eq:high_order_flux}.}
    \label{fig:high_order_stencil}
    \end{figure}

We make the observation that the high-order flux in Equation \ref{eq:high_order_flux} constructs the interface state as a weighted sum of multiple sub-interface contributions, where each sub-interface is formed from cell pairs $(m \mp \ell, m \mp \ell \pm i_s)$. This is shown clearly for orders 2, 4, and 6 in Figure \ref{fig:high_order_stencil}. For the internal energy flux, each sub-interface state can be expressed as:
\begin{equation}
\left.\rho e\right|^{\text{sub-interface}}_{(m \mp \ell,m \mp \ell \pm i_s)} = \frac{ \left. \rho e \right|_{(m \mp \ell)}+ \left. \rho e \right|_{\left(m \mp \ell \pm i_s\right)}}{2}
\end{equation}
which, when coupled with the corresponding velocity and density produces a pressure error analogous to the second-order case.

We therefore propose to apply an APEC correction to each sub-interface. By applying the correction derived in Equation \ref{eq:apec_rhoe_flux} to each cell pair in the stencil, we ensure that every sub-interface contribution maintains \textit{approximate} thermodynamic consistency between $\rho e$ and $\rho Y_i$. This leads to a high-order APEC scheme where the correction is distributed across all sub-interfaces weighted by their respective central difference coefficients $a_{q, i_s}$.

The high-order APEC corrected interface value for internal energy, $ \atfaceplus{\rho e}{m}$ is defined as 

\begin{equation}
\begin{aligned}
\label{eq:high_order_apec_rhoe_interface}
 \atfaceplus{\rho e}{m}^{\text{APEC $\mathcal{O}(q)$}} \equiv & 2 \sum_{i_s=1}^{q / 2} a_{q, i_s} \sum_{\ell=0}^{i_s-1} \frac{ \left .\rho e \right|_{(m \mp \ell)}+ \left. \rho e \right|_{\left(m \mp \ell \pm i_s\right)}}{2} \\
 & \quad - \sum_{i=1}^N \left[ 
 \frac{ \left. \left( \frac{\partial \rho e}{\partial \rho Y_i} \right)_{\rho Y_{j \neq i}, p} \right|_{(m \mp \ell)}-\left. \left( \frac{\partial \rho e}{\partial \rho Y_i} \right)_{\rho Y_{j \neq i}, p} \right|_{\left(m \mp \ell \pm i_s\right)}}{2} 
   \frac{ \left. \rho Y_{i} \right| _{(m \mp \ell)}- \left. \rho Y_{i} \right|_{\left(m \mp \ell \pm i_s\right)}}{2} 
  \right],
\end{aligned}
\end{equation}

where $N$ is the number of species,  $\rho Y_i$ is the mass density of species $k$. When defining the scheme, we also have the freedom to pick a flux or half-point values for $\rho Y_1$ and $\rho Y_2$ because both constraints were used in the derivation of the approximate condition (Equation \ref{eq:apec_rhoe_flux}). APEC chooses to make these the average of the neighboring points

\begin{equation}
\label{eq:rhoYi_interface_apec}    
\atfaceplus{\rho Y_i}{m}^\text{APEC} = \frac{\atcell{\rho Y_i}{m} + \atcell{\rho Y_i}{m+1}}{2}
\end{equation}

the equivalent in the high-order extension of APEC is to follow Equation \ref{eq:high_order_flux} which we will write explicitly for clarity

\begin{equation}
\label{eq:rhoYi_interface_high_order_apec}    
 \atfaceplus{\rho Y_i}{m} \equiv 2 \sum_{i_s=1}^{q / 2} a_{q, i_s} \sum_{\ell=0}^{i_s-1}  \frac{ \left. \rho Y_{i} \right|_{(m \mp \ell)}+ \left. \rho Y_{i} \right|_{\left(m \mp \ell \pm i_s\right)}}{2},
\end{equation}
where $i$ is the mass fraction number.

\subsection{Summary of Proposed Scheme - High-Order APEC}

Utilizing $\atfaceplus{\rho e}{m}$ as written in Equation \ref{eq:high_order_apec_rhoe_interface} we also define the interface of density as
\begin{equation}
\atfaceplus{\rho}{m}^{\text{APEC $\mathcal{O}(q)$}} = 2 \sum_{i_s=1}^{q / 2} a_{q, i_s} \sum_{\ell=0}^{i_s-1} \left[ \frac{ \left. \rho Y_{1} \right|_{(m \mp \ell)}+ \left. \rho Y_{1} \right|_{\left(m \mp \ell \pm i_s\right)}}{2}  + \frac{ \left. \rho Y_{2} \right|_{(m \mp \ell)}+ \left. \rho Y_{2} \right|_{\left(m \mp \ell \pm i_s\right)}}{2}\right].
\end{equation}

We can now write the complete high-order APEC scheme as


\begin{align}
    \atfaceplus{C_1}{m}^{\text{APEC $\mathcal{O}(q)$}} &= 2 \sum_{i_s=1}^{q / 2} a_{q, i_s} \sum_{\ell=0}^{i_s-1}  \frac{ \left. \rho Y_{1} \right|_{(m \mp \ell)}+ \left. \rho Y_{1} \right|_{\left(m \mp \ell \pm i_s\right)}}{2} \frac{ \left. u \right|_{(m \mp \ell)}+ \left. u \right|_{\left(m \mp \ell \pm i_s\right)}}{2} \\    
    \atfaceplus{C_2}{m}^{\text{APEC $\mathcal{O}(q)$}} &= 2 \sum_{i_s=1}^{q / 2} a_{q, i_s} \sum_{\ell=0}^{i_s-1}  \frac{ \left. \rho Y_{2} \right|_{(m \mp \ell)}+ \left. \rho Y_{2} \right|_{\left(m \mp \ell \pm i_s\right)}}{2} \frac{ \left. u \right|_{(m \mp \ell)}+ \left. u \right|_{\left(m \mp \ell \pm i_s\right)}}{2} \\
    \atfaceplus{M}{m}^{\text{APEC $\mathcal{O}(q)$}} &= 2\left( \atfaceplus{\rho}{m}^{\text{APEC $\mathcal{O}(q)$}} \right)  \sum_{i_s=1}^{q / 2} a_{q, i_s} \sum_{\ell=0}^{i_s-1}  \left[ \frac{ \left. u \right|_{(m \mp \ell)}+ \left. u \right|_{\left(m \mp \ell \pm i_s\right)}}{2} \right]^2\\
    \atfaceplus{\Pi}{m}^{\text{APEC $\mathcal{O}(q)$}} &= 2 \sum_{i_s=1}^{q / 2} a_{q, i_s} \sum_{\ell=0}^{i_s-1}  \frac{ \left. p \right|_{(m \mp \ell)}+ \left. p \right|_{\left(m \mp \ell \pm i_s\right)}}{2}   \\
    \atfaceplus{K}{m}^{\text{APEC $\mathcal{O}(q)$}} &= 2\left(\atfaceplus{\rho}{m}^{\text{APEC $\mathcal{O}(q)$}}\right) \sum_{i_s=1}^{q / 2} a_{q, i_s} \sum_{\ell=0}^{i_s-1}  \frac{ \left. u \right|_{(m \mp \ell)}+ \left. u \right|_{\left(m \mp \ell \pm i_s\right)}}{2} \frac{ \left. u \right|_{(m \mp \ell)} \left. u \right|_{\left(m \mp \ell \pm i_s\right)}}{2}\\
    \atfaceplus{I}{m}^{\text{APEC $\mathcal{O}(q)$}} &= 2\left( \atfaceplus{\rho e}{m}^{\text{APEC $\mathcal{O}(q)$}} \right) \sum_{i_s=1}^{q / 2} a_{q, i_s} \sum_{\ell=0}^{i_s-1}  \frac{ \left. u \right|_{(m \mp \ell)}+ \left. u \right|_{\left(m \mp \ell \pm i_s\right)}}{2} \\
    \atfaceplus{S}{m}^{\text{APEC $\mathcal{O}(q)$}} &= 2 \sum_{i_s=1}^{q / 2} a_{q, i_s}\sum_{\ell=0}^{i_s-1}  \frac{ \left. u \right|_{(m \mp \ell)} \left. p \right|_{\left(m \mp \ell \pm i_s\right)}+ \left. u \right|_{\left(m \mp \ell \pm i_s\right)} \left. p \right|_{(m \mp \ell)}}{2}
\end{align}

\subsection{Error Analysis of the High-Order APEC Scheme}

Since APEC is an approximate PEP scheme, it is worthwhile to analyze the error of pressure acquired by utilizing this scheme. We define a pressure error ``flux", following \cite{TERASHIMA2025113701}, as

\begin{equation}
\atcell{f_{PE}}{m} = \frac{\atfaceplus{\rho e}{m} - \atfaceminus{\rho e}{m}}{\Delta x} - \left.\sum_{i=1}^N\left(\frac{\partial \rho e}{\partial \rho Y_i}\right)_{\rho Y_{j \neq i}, p}\right|_m
\left(\frac{\left.\rho Y_i\right|_{m+\frac{1}{2}}-\left.\rho Y_i\right|_{m-\frac{1}{2}}}{\Delta x}\right),
\label{eq:pe_error_flux}
\end{equation}
which is Equation \ref{eq:pep_property_final}
rewritten to account for the fact that the scheme will not exactly satify this relationship and some pressure error will be present. For a PEP scheme $\atcell{f_{PE}}{m}$ should be exactly $0$. We can then choose the order of our method, in this work we'll focus on orders 2, 4, 6, and 8, and get the interface values needed to define the high-order APEC scheme, $\atfaceplus{\rho e}{m}$, $\atfaceplus{\rho Y_1}{m}$, and $\atfaceplus{\rho Y_2}{m}$ defined by Equations \ref{eq:high_order_apec_rhoe_interface} and \ref{eq:rhoYi_interface_high_order_apec} respectively. We perform a Taylor series analysis by expanding the variables $\rho Y_i$, $\rho e$, and the partial derivatives $\left(\frac{\partial \rho e}{\partial \rho Y_i}\right)_{\rho Y_{j \neq i}, p}$ about the cell center $m$ in terms of the grid spacing $\Delta x$. 

Each quantity $q$ at stencil point $m+\ell$ is represented as $q_{m+\ell} = \sum_{n=0}^{N_T} q^{(n)}_m (\ell \Delta x)^n / n!$, where $q^{(n)}_m$ denotes the $n^{\text{th}}$ derivative of $q$ evaluated at cell center $m$ and $N_T$ is chosen sufficiently large to capture the leading order terms. These Taylor expansions are then substituted into Equations \ref{eq:high_order_apec_rhoe_interface} and \ref{eq:rhoYi_interface_high_order_apec} to obtain symbolic expressions for the interface values $\atfaceplus{\rho e}{m}$, $\atfaceplus{\rho Y_i}{m}$ (and their corresponding minus-side values). Finally, these interface values are used to evaluate $\atcell{f_{PE}}{m}$ from Equation \ref{eq:pe_error_flux}, which is then expanded and we analyze the leading error term.

We note that the $0^{\text{th}}$ order term in all of the following taylor expansions is

\begin{equation}
\atcell{R_{PE}}{m} \equiv \atcell{\frac{\partial \rho e}{\partial x}}{m} - \sum_{i=0}^N \left( \atcell{\frac{\partial \rho e}{\partial \rho Y_i}}{m} \atcell{\frac{\partial \rho Y_i}{\partial x}}{m} \right)
\label{eq:R_PE_form}
\end{equation}

which we define as $\atcell{R_{PE}}{m}$ for the residual of pressure error at cell $m$. At timestep $n$ if the state is in pressure equilibrium then $\atcell{R_{PE}}{m}$ is exactly zero. However, as shown in Equation \ref{eq:time_inexact_pressure} even for a spatially exact PEP scheme, there will be some temporal error in pressure leading to the $0^{\text{th}}$ order propagation of $f_{PE}$ throughout the simulation through this $\atcell{R_{PE}}{m}$ term. Below, since everything is Taylor expanded around cell $m$ we drop the spacial subscripting notation for the remainder of this section.

\newcommand{\depsilon}[2]{%
  \frac{\partial^{#1} \left( \frac{\partial \rho e}{\partial \rho Y_{#2}} \right)}{\partial x^{#1}}}
\newcommand{\drhoY}[2]{\frac{\partial^{#1} \rho Y_{#2}}{\partial x^{#1}}}
\newcommand{\drhoe}[1]{\frac{\partial^{#1} \rho e}{\partial x^{#1}}}

\begin{align}
f_{PE}^{\text{KEEP $\mathcal{O}(2)$}} &= R_{PE}  + \frac{\Delta x^2}{6} \left[ \frac{\partial^3 \rho e}{\partial x^3} -\sum_{i=0}^{N} \left( \frac{\partial \rho e}{\partial \rho Y_i} \frac{\partial^3 \rho Y_i}{\partial x^3} \right)  \right]  +  \text{H.O.T.} \\
    f_{PE}^{\text{APEC $\mathcal{O}(2)$}} &= R_{PE}  + \frac{\Delta x^2}{6} \left[ \frac{\partial^3 \rho e}{\partial x^3} -\sum_{i=0}^{N} \left( \frac{\partial \rho e}{\partial \rho Y_i} \frac{\partial^3 \rho Y_i}{\partial x^3} \right)  \right]  \\ 
    &\quad - \frac{\Delta x^2}{4} \sum_{i=0}^{N}\left[\depsilon{}{i} \drhoY{2}{i} + \depsilon{2}{i} \drhoY{}{i} \right] \notag\\
    &\quad + \text{H.O.T.} \notag 
\end{align}

\begin{align}
f_{PE}^{\text{KEEP $\mathcal{O}(4)$}} &= R_{PE}  - \frac{\Delta x^4}{30} \left[ \frac{\partial^5 \rho e}{\partial x^5} -\sum_{i=0}^{N} \left( \frac{\partial \rho e}{\partial \rho Y_i} \frac{\partial^5 \rho Y_i}{\partial x^5} \right)  \right]  +  \text{H.O.T.} \\
f_{PE}^{\text{APEC $\mathcal{O}(4)$}} &= R_{PE}  - \frac{\Delta x^4}{30} \left[ \frac{\partial^5 \rho e}{\partial x^5} -\sum_{i=0}^{N} \left( \frac{\partial \rho e}{\partial \rho Y_i} \frac{\partial^5 \rho Y_i}{\partial x^5} \right)  \right]  \\ 
    &\quad + \frac{\Delta x^4}{12} \sum_{i=0}^{N} \left[
    \depsilon{}{i}\drhoY{4}{i} + 2\depsilon{2}{i} \drhoY{3}{i}
    + 2\depsilon{3}{i}\drhoY{2}{i} + \depsilon{4}{i}\drhoY{}{i}
    \right] \notag \\
    &\quad + \text{H.O.T.} \notag
\end{align}

\begin{align}
f_{PE}^{\text{KEEP $\mathcal{O}(6)$}} &= R_{PE}  + \frac{\Delta x^6}{140} \left[ \frac{\partial^7 \rho e}{\partial x^7} -\sum_{i=0}^{N} \left( \frac{\partial \rho e}{\partial \rho Y_i} \frac{\partial^7 \rho Y_i}{\partial x^7} \right)  \right]  +  \text{H.O.T.} \\
f_{PE}^{\text{APEC $\mathcal{O}(6)$}} &= R_{PE}  + \frac{\Delta x^6}{140} \left[ \frac{\partial^7 \rho e}{\partial x^7} -\sum_{i=0}^{N} \left( \frac{\partial \rho e}{\partial \rho Y_i} \frac{\partial^7 \rho Y_i}{\partial x^7} \right)  \right]  \\
&\quad- \frac{\Delta x^6}{40} \sum_{i=0}^{N} \left[ \depsilon{}{i}\drhoY{6}{i} + 3\depsilon{2}{i}\drhoY{5}{i} + 5 \depsilon{3}{i} \drhoY{4}{i} \right. \notag \\
&\quad + \left. 5\depsilon{4}{i} \drhoY{3}{i} + 3\depsilon{5}{i} \drhoY{2}{i} + \depsilon{6}{i}\drhoY{}{i} \right] \notag\\
&\quad +\text{H.O.T.} \notag 
\end{align}

\begin{align}
f_{PE}^{\text{KEEP $\mathcal{O}(8)$}} &= R_{PE}  - \frac{\Delta x^8}{630} \left[ \frac{\partial^9 \rho e}{\partial x^9} -\sum_{i=0}^{N} \left( \frac{\partial \rho e}{\partial \rho Y_i} \frac{\partial^9 \rho Y_i}{\partial x^9} \right)  \right]  +  \text{H.O.T.} \\
f_{PE}^{\text{APEC $\mathcal{O}(8)$}} &= R_{PE}  - \frac{\Delta x^8}{630} \left[ \frac{\partial^9 \rho e}{\partial x^9} -\sum_{i=0}^{N} \left( \frac{\partial \rho e}{\partial \rho Y_i} \frac{\partial^9 \rho Y_i}{\partial x^9} \right)  \right]  \\
&\quad + \frac{\Delta x^8}{630} \sum_{i=0}^{N} \left[  \frac{9}{2} \depsilon{1}{i} \drhoY{8}{i} + 18\depsilon{2}{i}\drhoY{7}{i} + 42 \depsilon{3}{i} \drhoY{6}{i} + 63 \depsilon{4}{i} \drhoY{5}{i}\right. \notag \\
&\quad + \left. 63 \depsilon{5}{i}\drhoY{4}{i} + 42 \depsilon{6}{i} \drhoY{3}{i} + 18 \depsilon{7}{i}\drhoY{2}{i} + \frac{9}{2}\depsilon{8}{i}\drhoY{1}{i} \right] \notag \\
&\quad +\text{H.O.T.} \notag
\end{align}

In order to simplify these statements, we let the current state at timestep $n$ to be in pressure equilibrium which sets $R_{PE} = 0$. This focuses the analysis on the accumulation of pressure error per timestep rather then the propagation of already accumulated pressure error. This gives us the ability to simplify further by noting that one can take the analytical $n^{\text{th}}$ derivative of $\rho e$ with respect to $x$ in Equation \ref{eq:R_PE_form} via the chain rule which can be expanded as

\begin{equation}
    \label{eq:leibniz_rule_rhoe}
    \drhoe{n} =  \sum_{i=1}^N \left( \sum_{k=0}^q \binom{q}{k} \frac{d^k}{dx^k}\left(\frac{\partial \rho e}{\partial(\rho Y_i)}\right) \frac{d^{n-k}}{dx^{n-k}}\left(\frac{\partial(\rho Y_i)}{\partial x}\right) \right),
\end{equation}

where $q$ is the order of the scheme, $N$ is the number of species, and $n$ is the order of derivative. We can use Equation \ref{eq:leibniz_rule_rhoe} to convert the terms from the APEC Taylor expansion and the KEEP taylor expansion to compare them directly. For orders 2, 4, 6, and 8 we simplify to the following:

\begin{equation}
\label{eq:simplified_keep_2}
    f_{PE}^{\text{KEEP $\mathcal{O}(2)$}} = \sum_{i=1}^N \left[ \frac{1}{3 }\depsilon{1}{i} \drhoY{2}{i} + \frac{1}{6}\depsilon{2}{i} \drhoY{1}{i} \right],
\end{equation}
\begin{equation}
\label{eq:simplified_apec_2}
    f_{PE}^{\text{APEC $\mathcal{O}(2)$}} = \sum_{i=1}^N \left[ \frac{1}{12}\depsilon{1}{i} \drhoY{2}{i} - \frac{1}{12}\depsilon{2}{i} \drhoY{1}{i} \right],
\end{equation}

\begin{equation}
\label{eq:simplified_keep_4}
    f_{PE}^{\text{KEEP $\mathcal{O}(4)$}} = \sum_{i=1}^N \left[-\frac{2}{15} \depsilon{1}{i} \drhoY{4}{i} - \frac{1}{5} \depsilon{2}{i} \drhoY{3}{i} - \frac{2}{15} \depsilon{3}{i} \drhoY{2}{i} -\frac{1}{30} \depsilon{4}{i} \drhoY{1}{i} \right],
\end{equation}
\begin{equation}
\label{eq:simplified_apec_4}
    f_{PE}^{\text{APEC $\mathcal{O}(4)$}} = \sum_{i=1}^N \left[-\frac{1}{20} \depsilon{1}{i} \drhoY{4}{i} - \frac{1}{30} \depsilon{2}{i} \drhoY{3}{i} + \frac{1}{30} \depsilon{3}{i} \drhoY{2}{i} +\frac{1}{20} \depsilon{4}{i} \drhoY{1}{i} \right],
\end{equation}

\begin{equation}
\label{eq:simplified_keep_6}
\begin{split}
    f_{PE}^{\text{KEEP $\mathcal{O}(6)$}} &= \sum_{i=1}^N \left[\frac{3}{70} \depsilon{1}{i} \drhoY{6}{i} + \frac{3}{28} \depsilon{2}{i} \drhoY{5}{i} + \frac{1}{7}\depsilon{3}{i} \drhoY{4}{i}\right. \\ &\quad+ \left.\frac{3}{28} \depsilon{4}{i} \drhoY{3}{i} + \frac{3}{70} \depsilon{5}{i} \drhoY{2}{i} + \frac{1}{140} \depsilon{6}{i} \drhoY{1}{i}\right],
\end{split}
\end{equation}
\begin{equation}
\begin{split}
\label{eq:simplified_apec_6}
    f_{PE}^{\text{APEC $\mathcal{O}(6)$}} &= \sum_{i=1}^N \left[\frac{1}{56} \depsilon{1}{i} \drhoY{6}{i} + \frac{9}{280} \depsilon{2}{i} \drhoY{5}{i} + \frac{1}{56}\depsilon{3}{i} \drhoY{4}{i}\right. \\ &\quad - \left. \frac{1}{56} \depsilon{4}{i} \drhoY{3}{i} - \frac{9}{280} \depsilon{5}{i} \drhoY{2}{i} - \frac{1}{56} \depsilon{6}{i} \drhoY{1}{i} \right],
\end{split}
\end{equation}

\begin{equation}
\begin{split}
\label{eq:simplified_keep_8}
    f_{PE}^{\text{KEEP $\mathcal{O}(8)$}} &= \sum_{i=1}^N \left[- \frac{4}{315} \depsilon{1}{i} \drhoY{8}{i} -\frac{2}{45} \depsilon{2}{i} \drhoY{7}{i} - \frac{4}{45}\depsilon{3}{i} \drhoY{6}{i} -\frac{1}{9} \depsilon{4}{i} \drhoY{5}{i} \right.\\ 
    & \quad \left. - \frac{4}{45} \depsilon{5}{i} \drhoY{4}{i} - \frac{2}{45} \depsilon{6}{i} \drhoY{3}{i} - \frac{4}{315} \depsilon{7}{i} \drhoY{2}{i}  - \frac{1}{630} \depsilon{8}{i} \drhoY{1}{i} \right],
\end{split}
\end{equation}

\begin{equation}
\begin{split}
\label{eq:simplified_apec_8}
    f_{PE}^{\text{APEC $\mathcal{O}(8)$}} &= \sum_{i=1}^N \left[- \frac{1}{180} \depsilon{1}{i} \drhoY{8}{i} -\frac{1}{63} \depsilon{2}{i} \drhoY{7}{i} - \frac{1}{45}\depsilon{3}{i} \drhoY{6}{i} -\frac{1}{90} \depsilon{4}{i} \drhoY{5}{i} \right.\\ 
    & \quad \left.+ \frac{1}{90} \depsilon{5}{i} \drhoY{4}{i} + \frac{1}{45} \depsilon{6}{i} \drhoY{3}{i} + \frac{1}{63} \depsilon{7}{i} \drhoY{2}{i}  + \frac{1}{180} \depsilon{8}{i} \drhoY{1}{i}\right].
\end{split}
\end{equation}

The Taylor expansions show that, at a specified order $q$, the high-order APEC construction yields uniformly smaller leading coefficients in the pressure-error flux $f_{PE}$ than the corresponding KEEP discretization, together with a sign structure that promotes cancellation among mixed derivatives. In particular,
comparing the second-, fourth-, sixth-, and eighth-order KEEP forms in
Equation~\ref{eq:simplified_keep_2}, Equation~\ref{eq:simplified_keep_4},
Equation~\ref{eq:simplified_keep_6}, and Equation~\ref{eq:simplified_keep_8}
with their APEC counterparts in Equation~\ref{eq:simplified_apec_2},
Equation~\ref{eq:simplified_apec_4}, Equation~\ref{eq:simplified_apec_6}, and
Equation~\ref{eq:simplified_apec_8} shows strictly smaller magnitude prefactors
multiplying the mixed products \(\depsilon{n}{i}\,\drhoY{m}{i}\) and, for APEC,
alternating signs that induce partial cancellation across the stencil. There
are three isolated exceptions to the strictly smaller prefactors---two
composition-derivative terms in the \(\mathcal{O}(8)\) scheme and one
first-derivative composition term in the \(\mathcal{O}(6)\) scheme---but these
constitute a small subset of the many contributing coefficients and do not
alter the overall trend that APEC systematically reduces the pressure-error
contribution relative to KEEP.

For a given mesh and $q$, the high-order APEC scheme preserves the formal order of the underlying central operator while delivering smaller truncation constants in $f_{PE}$, implying reduced spurious pressure oscillations and improved near-equilibrium preservation under arbitrary EOS, while temporal propagation through $R_{PE}$ proceeds as in the baseline formulation.

\section{Derivation of a Fully Conservative PEP Scheme}
\label{sec:fully_conservative_PEP}

We now begin the derivation of a fully conservative, spatially exact PEP scheme. Such a scheme must satisfy Equations \ref{eq:exact_condition_right} and \ref{eq:exact_condition_left_shifted} exactly. Here we treat these as separate constraints, rather than combining them into one as done in APEC, which yields two equations for three unknown interface quantities: $\atfaceplus{\rho Y_1}{m}$, $\atfaceplus{\rho Y_2}{m}$, and $\atfaceplus{\rho e}{m}$. This underdetermined system admits multiple possible closures, each corresponding to a distinct conservative PEP variant. 

\subsection{An Example PEP Scheme}
\label{sec:simple_example_pep}

The simplest possible variant of this is to let $\atfaceplus{\rho e}{m}$ be equal to the average values of the cells adjacent via

\begin{equation}
\label{eq:halfpoint_rhoy1}
\atfaceplus{\rho e}{m} = \frac{\atcell{\rho e}{m} + \atcell{\rho e}{m+1}}{2}.
\end{equation}

With this choice of  $\rho e |_{m + \frac{1}{2}}$, we are left with a $2\times2$ system to solve. We define 

\begin{equation}
\epsilon_i \equiv \left( \frac{\partial \rho e}{\partial \rho Y_i} \right)_{\rho Y_{j \neq i}, p}
\end{equation}

and note that $\rho e |_{m + \frac{1}{2}}$ is defined in some arbitrary manner and it can be considered a known quantity. We can then write the system as

\begin{equation}
    \begin{pmatrix}
        \epsilon_1 |_{m+1} & \epsilon_2 |_{m+1} \\[4pt]
        \epsilon_1 |_{m} & \epsilon_2 |_{m}
    \end{pmatrix} 
    \begin{pmatrix}
        \rho Y_1 |_{m+\frac{1}{2}} \\[4pt]
        \rho Y_2 |_{m+\frac{1}{2}}
    \end{pmatrix} 
    = 
    \begin{pmatrix}
        \rho e |_{m + \frac{1}{2}} - \rho e |_{m+1} + \epsilon_1 |_{m+1}\rho Y_1|_{m+1} + \epsilon_2 |_{m+1}\rho Y_2|_{m+1} \\[4pt]
        \rho e |_{m + \frac{1}{2}} - \rho e |_{m} + \epsilon_1 |_{m}\rho Y_1|_{m} + \epsilon_2 |_{m}\rho Y_2|_{m}  
    \end{pmatrix}. 
    \label{eq:pep_matrix}
\end{equation}

As $\epsilon_i|_{m+1} \to \epsilon_i|_{m}$, the two rows of the coefficient matrix in Equation \ref{eq:pep_matrix} become nearly linearly dependent and its determinant tends to zero. In this regime the system is increasingly ill-conditioned. An equivalent conclusion is obtained if one treats a mass fraction as a free variable and instead solves for $\rho Y_1|_{m+\frac{1}{2}}$ and $\rho e|_{m+\frac{1}{2}}$. We therefore expect the linear system Equation \ref{eq:pep_matrix} to be numerically unreliable when the left and right states are nearly identical.

\subsection{Equal/near-equal states.}
\label{sec:equal_states}
As we have seen, as $|\epsilon_i|_{m+1} - \epsilon_i|_m| \to 0$ the PEP scheme becomes unstable, but an important realization is that this is exactly where the scheme is not needed.
Assume uniform pressure and $\epsilon_i|_{m}=\epsilon_i|_{m+1}=\epsilon_i^\star$ we then rewrite Equations \ref{eq:exact_condition_right} and \ref{eq:exact_condition_left_shifted} as
\begin{align}
\left.\rho e\right|_{m+\frac{1}{2}}-\left.\rho e\right|_{m}
&= \sum_{i=1}^N \epsilon_i^\star\Big(\left.\rho Y_i\right|_{m+\frac{1}{2}}-\left.\rho Y_i\right|_{m}\Big),\\
\left.\rho e\right|_{m+1}-\left.\rho e\right|_{m+\frac{1}{2}}
&= \sum_{i=1}^N \epsilon_i^\star\Big(\left.\rho Y_i\right|_{m+1}-\left.\rho Y_i\right|_{m+\frac{1}{2}}\Big).
\end{align}
Adding the two gives the cell-centered relation
\begin{equation}
\left.\rho e\right|_{m+1}-\left.\rho e\right|_{m}
= \sum_{i=1}^N \epsilon_i^\star\Big(\left.\rho Y_i\right|_{m+1}-\left.\rho Y_i\right|_{m}\Big),
\label{eq:cell_identity_short}
\end{equation}
so the two interface constraints are consistent with the single condition at cell centers. When $\epsilon_i|_{m}=\epsilon_i|_{m+1}$, this cell-centered relation is exact, and the centered averages
\begin{equation}
\left.\rho e\right|_{m+\frac{1}{2}}=\tfrac{1}{2}\big(\left.\rho e\right|_{m}+\left.\rho e\right|_{m+1}\big),\qquad
\left.\rho Y_i\right|_{m+\frac{1}{2}}=\tfrac{1}{2}\big(\left.\rho Y_i\right|_{m}+\left.\rho Y_i\right|_{m+1}\big)
\end{equation}
satisfy the pressure–equilibrium condition at each interface exactly (Equations \ref{eq:exact_condition_right} and \ref{eq:exact_condition_left_shifted}), since both expressions reduce to one half of the corresponding cell-centered difference

\subsection{Proposed PEP Scheme}
\label{sec:proposed_scheme}
In the design of the proposed PEP scheme, we keep in mind that the PEP linear system, Equation \ref{eq:pep_matrix}, becomes ill-conditioned in regions where $|\epsilon_i|_{m+1} \approx |\epsilon_i|_{m}$—precisely where strong composition jumps are absent and a PEP correction is not essential. We therefore make the following practical decision: rather than solving directly for an unknown interface value $\atfaceplus{\phi}{m}$, we choose to assume the form
\begin{equation}
    \label{eq:general_form_eq}
    \atfaceplus{\phi}{m} = \frac{\atcell{\phi}{m} + \atcell{\phi}{m+1}}{2} + C
\end{equation}

where $C$ is now the unknown constant we are solving. This has two benefits. 
First, we know by the discussion in Section \ref{sec:equal_states} that the simple average half point values satisfy the PEP criterion in the limit that the left state equals the right state. This gives us the additional information that $C$ should approach zero (which we will prove later) in this limit. This is easier to satisfy than the limiting behavior in \ref{sec:simple_example_pep} which would need to approach the half point average in the near equal limit. We are simply isolating all the numerical instability into $C$ which can easily be set to $0$ in regions where we detect this is needed. The second benefit to this is we can modify any numerical scheme to be pressure equilibrium preserving. When these coefficients are zero, we will exactly recover the underlying numerical scheme of our choosing. With the goal of pressure equilibrium, we let this underlying method be APEC \cite{TERASHIMA2025113701}. We will now present the proposed half point values of the scheme

\begin{equation}
\label{eq:halfpoint_rhoy1_proposed}
\atfaceplus{\rho Y_1}{m}^\text{PEP} = \frac{\atcell{\rho Y_1}{m} + \atcell{\rho Y_1}{m+1}}{2}  + \alpha,
\end{equation}

\begin{equation}
\label{eq:halfpoint_rhoy2_proposed}
\atfaceplus{\rho Y_2}{m}^\text{PEP} = \frac{\atcell{\rho Y_2}{m} + \atcell{\rho Y_2}{m+1}}{2}  - \alpha,
\end{equation}

\begin{equation}
\label{eq:halfpoint_rhoe_proposed}
\atfaceplus{\rho e}{m}^\text{PEP} = \frac{\atcell{\rho e}{m} + \atcell{\rho e}{m+1}}{2}  + \beta  - \sum_{i=1}^N{} \left[ \frac{\epsilon_i |_{m+1}}{2} \left(\rho Y_i |_{m+1} - \rho Y_i |_{m+\frac{1}{2}} \right) - \frac{\epsilon_i |_{m}}{2} \left(\rho Y_i |_{m+\frac{1}{2}} - \rho Y_i |_{m} \right) \right].
\end{equation}

For consistency between mass fraction we choose to modify  $\rho e |_{m + \frac{1}{2}}$ as well. This scheme also maintains the flux of $\rho$ exactly as the underlying scheme, in this case the APEC scheme, unlike \cite{fujiwara_fully_2023}. This decomposition of the flux has a very useful feature in analyzing the determinate limit of the formulation because it is now isolated only in $\alpha$ and $\beta$, which we only need to prove that their limit is zero for consistency. Furthermore, this decomposition allows for the most simple treatment of the near-limiting behavior, where floating point error can pollute the calculation of the fraction where both numerator and denominator are near zero. The key to this simple treatment is that the choice of the underlying scheme is arbitrary, as long as it represents a consistent flux.

We now have two unknowns, $\alpha$ and $\beta$, paired with our two constraints, Equations \ref{eq:exact_condition_right} and \ref{eq:exact_condition_left_shifted}. We can solve these analytically and write the algebraic closed form as:



\begin{equation}
\label{eq:alpha}
\alpha =
\frac{
\displaystyle
\sum_{i=1}^N
\bigl(\atcell{\epsilon_i}{m} + \atcell{\epsilon_i}{m+1}\bigr)
\bigl(\atcell{\rho Y_i}{m} - \atcell{\rho Y_i}{m+1}\bigr)
+ 2\bigl(\atcell{\rho e}{m+1} - \atcell{\rho e}{m}\bigr)
}{
2\Big[
\bigl(\atcell{\epsilon_1}{m} - \atcell{\epsilon_1}{m+1}\bigr)
-
\bigl(\atcell{\epsilon_2}{m} - \atcell{\epsilon_2}{m+1}\bigr)
\Big]
},
\end{equation}

\begin{equation}
\label{eq:beta}
\beta =
\frac{
\Big[
\big(\atcell{\epsilon_1}{m} + \atcell{\epsilon_1}{m+1}\big)
-
\big(\atcell{\epsilon_2}{m} + \atcell{\epsilon_2}{m+1}\big)
\Big]
\left[
\displaystyle
\sum_{i=1}^N
\big(\atcell{\epsilon_i}{m} + \atcell{\epsilon_i}{m+1}\big)
\big(\atcell{\rho Y_i}{m} - \atcell{\rho Y_i}{m+1}\big)
- 2\big(\atcell{\rho e}{m} - \atcell{\rho e}{m+1}\big)
\right]
}{
4\Big[
\big(\atcell{\epsilon_1}{m} - \atcell{\epsilon_1}{m+1}\big)
-
\big(\atcell{\epsilon_2}{m} - \atcell{\epsilon_2}{m+1}\big)
\Big]
}.
\end{equation}

As expected, $\alpha$ and $\beta$ inherit the same numerical instability problem because their denominator will approach 0 as $\atcell{\epsilon_i}{m} \approx\atcell{\epsilon_i}{m+1}$. From here there are two equivalent approaches, one could design some switching where based on some tolerance based on $\left| \atcell{\epsilon_i}{m+1} - \atcell{\epsilon_i}{m}\right|$ one can either choose to evaluate $\alpha,\beta$ algebraically if it is numerically safe or let $\alpha,\beta = 0$ and recover the second order APEC scheme. The second approach and the one we will take in this work is we retain the system of equations but rewrite such that we solve for $\alpha,\beta$ directly

\begin{equation}
\begin{aligned}
&\vcenter{\hbox{$
\begin{bmatrix}
\atcell{\epsilon_1}{m+1} - \atcell{\epsilon_2}{m+1} & -1 \\[4pt]
-\atcell{\epsilon_1}{m} + \atcell{\epsilon_2}{m} & 1
\end{bmatrix}
\begin{bmatrix}
\alpha \\[4pt] \beta
\end{bmatrix}
$}} =
\\[12pt]
&
\begin{bmatrix}
\displaystyle
\sum_i^N \Bigg[
\atcell{\epsilon_i}{m+1}
\left(-\tfrac{1}{2}\atcell{\rho Y_i}{m}
+ \tfrac{1}{2}\atcell{\rho Y_i}{m+1}\right)
- \left(-\tfrac{\atcell{\epsilon_i}{m}}{2}
+ \tfrac{\atcell{\epsilon_i}{m+1}}{2}\right)
\frac{-\atcell{\rho Y_i}{m} + \atcell{\rho Y_i}{m+1}}{2}
\Bigg]
+ \tfrac{1}{2}\left(\atcell{\rho e}{m} - \atcell{\rho e}{m+1}\right)
\\[12pt]
\displaystyle
\sum_i^N \Bigg[
\atcell{\epsilon_i}{m}
\left(-\tfrac{1}{2}\atcell{\rho Y_i}{m}
+ \tfrac{1}{2}\atcell{\rho Y_i}{m+1}\right)
+ \left(-\tfrac{\atcell{\epsilon_i}{m}}{2}
+ \tfrac{\atcell{\epsilon_i}{m+1}}{2}\right)
\frac{-\atcell{\rho Y_i}{m} + \atcell{\rho Y_i}{m+1}}{2}
\Bigg]
+ \tfrac{1}{2}\left(\atcell{\rho e}{m} - \atcell{\rho e}{m+1}\right)
\end{bmatrix}.
\end{aligned}
\label{eq:system}
\end{equation}

Solving this system requires using a pseudo-inverse algorithm for numerical stability. We leverage the Moore-Penrose \cite{moore1920reciprocal} algorithm which has a reciprocal condition number parameter. This is a tolerance parameter used to determine which singular values should be treated as effectively zero. For the purposes of the PEP scheme, this parameter controls how smoothly we transition between the fully conservative PEP update and the approximate APEC flux in space. Larger values produce more gradual spatial changes in the flux formulation, which can improve robustness by avoiding abrupt local switches but also enlarge the region where the lower-accuracy APEC approximation is active; smaller values sharpen the transition, preserving the high-order PEP formulation over a wider portion of the domain but making the scheme more sensitive to roundoff and localized instability. Over-utilizing the APEC correction can therefore introduce an unnecessary loss of accuracy, while under-utilizing it risks allowing small pressure imbalances to be generated from numerical precision issues in computing the PEP correction terms. In practice, we bias this parameter toward the conservative side, since the scheme still performs well with moderate APEC utilization and long-term stability of the pressure field is a key metric in this work. There is some flexibility to this choice, but through some analysis we can derive a \emph{safe} choice which we will utilize.

\subsection{Derivation of a Global Reciprocal Condition Number Estimate}
We now turn to deriving a single global reciprocal condition number threshold, which we will denote $r_g$, to be used in the Moore--Penrose pseudoinverse for the local $(\alpha,\beta)$ solve at every interface. We base $r_g$ on a scale-robust conditioning diagnostic of the underlying $2\times2$ system, since for non-ideal EOS the raw condition number can be dominated by the magnitude of $\theta=\epsilon_1-\epsilon_2$ and appear large even when the system is not actually close to the singular limit $\theta_{m+1}\approx\theta_m$.

In order to derive $r_g$ we look at the conditioning of the local $2\times 2$ system used to compute $(\alpha,\beta)$. At each cell we form
\begin{equation}
\atcell{\theta}{m} \;\equiv\; \atcell{\epsilon_{1}}{m}-\atcell{\epsilon_2}{m},
\end{equation}
since $\theta$ is the coefficient that controls when the two interface constraints become nearly redundant in Equation \ref{eq:system}. Because the magnitude of $|\theta|$ can vary widely across EOS and thermodynamic states, we normalize the $\theta$-column only for the conditioning estimate by introducing

\begin{equation}
\atfaceplus{s}{m}=\sqrt{\max\!\left(|\atcell{\theta}{m}|,|\atcell{\theta}{m+1}|,1\right)},\qquad
\atfaceplus{\widetilde{A}}{m}=
\begin{bmatrix}
\atcell{\theta}{m+1}/\atfaceplus{s}{m} & -1\\
-\atcell{\theta}{m}/\atfaceplus{s}{m} & 1
\end{bmatrix}.
\end{equation}

We then define the symmetric matrix
\begin{equation}
\label{eq:g_sym}
\atfaceplus{G}{m}\;\equiv\;\atfaceplus{\widetilde{A}^{\top}}{m} \atfaceplus{\widetilde{A}}{m},
\end{equation}

whose eigenvalues $0\le\lambda_{\min}\le\lambda_{\max}$ provide a direct measure of conditioning. If $\atfaceplus{\widetilde{A}}{m}$ is singular then $\lambda_{\min}=0$, and if it is nearly singular then $\lambda_{\min}\ll \lambda_{\max}$. In particular, the condition number of $\atfaceplus{G}{m}$ can be written as
\begin{equation}
 \kappa(\atfaceplus{G}{m})
\;=\;\frac{\lambda_{\max}(\atfaceplus{G}{m})}{\lambda_{\min}(\atfaceplus{G}{m})},
\end{equation}
and we define the local \emph{reciprocal} condition number
\begin{equation}
\atfaceplus{r}{m}
\;=\;\frac{\lambda_{\min}(\atfaceplus{G}{m})}{\lambda_{\max}(\atfaceplus{G}{m})},
\end{equation}
which decreases toward $0$ as the local solve becomes more ill-conditioned. In order to choose a single global tolerance, we simply take the maximum over all faces at the initial condition
\begin{equation}
r_g \;=\; \max \left(10^{-13}, \max_{m \in Nx}\; \atfaceplus{r}{m} \right),
\label{eq:rcond_global}
\end{equation}

where we apply a small lower bound $10^{-13}$ to avoid effectively vanishing tolerances in nearly singular regions. This value is then used in the Moore--Penrose pseudoinverse when solving the original (unscaled) system $A(\alpha,\beta)^{\top}=b$ at every interface. In our implementation, $r_g$ is computed once from the initial condition and held fixed for the remainder of the simulation. Taking the maximum over faces, together with the lower bound, provides a simple, robust choice: it admits slightly more pressure error than strictly necessary in some regions, but avoids introducing additional local switching and performs well for all test cases considered here. For more complex problems, one could instead recompute $r$ periodically (e.g.\ each timestep) or allow a space-dependent $\atfaceplus{r}{m}$ to better balance accuracy and robustness. For additional information on the sensitivity to this parameter and values used in this study, we point the reader to \ref{sec:rcond_sensitivity}.

\subsection{Extension to Multiple Dimensions and N Species}
\label{sub:nspecies}
The construction of the proposed PEP scheme as well as the high-order APEC scheme generalizes directly to $d$ spatial dimensions and $N$ species. Because the correction acts only on the mass fraction and total–energy fluxes, the momentum flux is advanced with the standard multidimensional discretization and no additional terms are required. Extending the scheme to $N$ species requires modifying only two species to enforce the criterion in Equations  \ref{eq:exact_condition_right} and \ref{eq:exact_condition_left_shifted} while leaving all remaining components unchanged. The choice of which two to correct is flexible, though we recommend selecting the two species with the largest mass fractions.

\subsection{Summary of Proposed PEP Scheme}
We summarize the proposed PEP scheme as a modification to the KEEP scheme where $\atfaceplus{\rho Y_1}{m}^\text{PEP}$, $\atfaceplus{\rho Y_1}{m}^\text{PEP}$ and $\atfaceplus{\rho e}{m}^\text{PEP}$ are defined in Equations \ref{eq:halfpoint_rhoy1_proposed}, \ref{eq:halfpoint_rhoy2_proposed}, and \ref{eq:halfpoint_rhoe_proposed} respectively. We let $ \atfaceplus{\rho}{m}^\text{PEP} = \atfaceplus{\rho Y_1}{m}^\text{PEP} +\atfaceplus{\rho Y_2}{m}^\text{PEP}$.

\begin{align}
    \atfaceplus{C_1}{m} &= \atfaceplus{\rho Y_1}{m}^\text{PEP} \frac{\atcell{u}{m} + \atcell{u}{m+1}}{2}     \label{eq:proposed_pep_beg} \\
    \atfaceplus{C_2}{m} &= \atfaceplus{\rho Y_2}{m}^\text{PEP} \frac{\atcell{u}{m} + \atcell{u}{m+1}}{2} \\
    \atfaceplus{M}{m} &= \atfaceplus{\rho}{m}^\text{PEP} \frac{\atcell{u}{m} + \atcell{u}{m+1}}{2} \frac{\atcell{u}{m} + \atcell{u}{m+1}}{2} \\
    \atfaceplus{\Pi}{m} &= \frac{\atcell{p}{m} + \atcell{p}{m+1}}{2} \\
    \atfaceplus{K}{m} &= \atfaceplus{\rho}{m} \frac{\atcell{u}{m} + \atcell{u}{m+1}}{2} \frac{\atcell{u}{m} \atcell{u}{m+1}}{2} \\
    \atfaceplus{I}{m} &= \atfaceplus{\rho e}{m}^\text{PEP} \frac{\atcell{u}{m} + \atcell{u}{m+1}}{2} \\
    \atfaceplus{S}{m} &= \frac{\atcell{u}{m} \atcell{p}{m+1} + \atcell{u}{m+1} \atcell{p}{m}}{2}
    \label{eq:proposed_pep_end}
\end{align}

\subsection{Limiting behavior and consistency of the scheme}
We now consider the limiting behavior of $\alpha$ and $\beta$ as shown in Equation \ref{eq:alpha} and \ref{eq:beta}. The goal is to show that these quantities approach 0 as the left and right state are equal to each other. This will then recover the original APEC scheme in the near equal limit and the scheme will be consistent. We analyze the numerator and denominator of each coefficient separately. 

\paragraph{Numerator of $\alpha$} Starting with the numerator of $\alpha$ we define 

\begin{equation}
\text{num}_\alpha = \sum_i^N \left[ \left( \atcell{\epsilon_i}{m} + \atcell{\epsilon_i}{m+1}\right)
    \left(\atcell{\rho Y_i}{m} - \atcell{\rho Y_i}{m + 1}\right)\right] + 2 \left( \atcell{\rho e}{m+1} - \atcell{\rho e }{m} \right).
\label{eq:num_alpha}
\end{equation}

We also define $\delta_i \equiv \atcell{\rho Y_i}{m +1} - \atcell{\rho Y_i}{m}$ and we Taylor expand $\epsilon$ at cell $m+1$ around cell $m$ as 

\begin{equation}
\atcell{\epsilon_i}{m+1} =\atcell{\epsilon_i}{m} + \sum_i^N \atcell{\xi_{i,j}}{m} \delta_j + O(\delta^2).
\label{eq:epsilon_taylor_expand}
\end{equation}

Similarly we expand $\atcell{\rho e}{m+1 }$ as

\begin{equation}
    \atcell{\rho e}{m+1 } = \atcell{\rho e}{m} + \sum_j \atcell{\epsilon_j}{m} \delta_j + O(\delta^2).
\label{eq:rhoe_taylor_expand}
\end{equation}

Plugging Equation \ref{eq:epsilon_taylor_expand} and \ref{eq:rhoe_taylor_expand} into Equation \ref{eq:num_alpha} we obtain

\begin{equation}
\text{num}_\alpha = \sum_i - 2 \atcell{\epsilon}{m} \delta_i - \sum_j\atcell{\xi_j}{m}\delta_j^2 + \sum_i 2\atcell{\epsilon}{m}\delta_i.
\label{eq:num_alpha_taylor_expand}
\end{equation}

Which after simplification implies $\text{num}_\alpha \propto O(\delta^2) $. 





\paragraph{Numerator of $\beta$} We analyze $\beta$ the same way where we define

\begin{align}
    \text{num}_\beta
    &=
    \Big[
      \big(\atcell{\epsilon_1}{m} + \atcell{\epsilon_1}{m+1}\big)
      -
      \big(\atcell{\epsilon_2}{m} + \atcell{\epsilon_2}{m+1}\big)
    \Big]
    \Bigg(
      \sum_{i=1}^N
      \big(\atcell{\epsilon_i}{m} + \atcell{\epsilon_i}{m+1}\big)
      \big(\atcell{\rho Y_i}{m} - \atcell{\rho Y_i}{m+1}\big)
      - 2\big(\atcell{\rho e}{m} - \atcell{\rho e}{m+1}\big)
    \Bigg).
    \label{eq:num_beta}
\end{align}
We again use the Equations
\ref{eq:epsilon_taylor_expand} and \ref{eq:rhoe_taylor_expand} and 
substitute these into \ref{eq:num_beta} and collecting like terms, all
contributions at order $\mathcal{O}(\delta)$ cancel:
\begin{equation}
    \text{num}_\beta
    =
    -\,2\big(\atcell{\epsilon_1}{m} - \atcell{\epsilon_2}{m}\big)
    \sum_{i,j} \atcell{\xi_{i,j}}{m}\,\delta_i \delta_j
    + \mathcal{O}(\delta^3).
    \label{eq:num_beta_taylor_expand}
\end{equation}
This implies $\text{num}_\beta \propto \mathcal{O}(\delta^2)$.

\paragraph{Denominator of $\beta$}
For the denominator of $\beta$ we similarly define
\begin{equation}
    \text{denom}_\beta
    =
    4\Big[
    \big(\atcell{\epsilon_1}{m} - \atcell{\epsilon_1}{m+1}\big)
    -
    \big(\atcell{\epsilon_2}{m} - \atcell{\epsilon_2}{m+1}\big)
    \Big].
    \label{eq:denom_beta}
\end{equation}
Using the same Taylor expansion shown in Equation \ref{eq:epsilon_taylor_expand} for
$\atcell{\epsilon_i}{m+1}$ and expressing the result in terms of $\delta_j$,
we obtain
\begin{equation}
\begin{aligned}
    \text{denom}_\beta
    &=
    4\Big[
    -\sum_{j}\atcell{\xi_{1,j}}{m}\delta_j
    +
    \sum_{j}\atcell{\xi_{2,j}}{m}\delta_j
    \Big]
    + \mathcal{O}(\delta^2) \\
    &=
    -\,4\sum_{j}
    \big(\atcell{\xi_{1,j}}{m} - \atcell{\xi_{2,j}}{m}\big)\,\delta_j
    + \mathcal{O}(\delta^2).
    \label{eq:denom_beta_taylor_expand}
\end{aligned}
\end{equation}
This then implies $\text{denom}_\beta \propto \mathcal{O}(\delta)$.
Combining Equations \ref{eq:num_beta_taylor_expand} and
\ref{eq:denom_beta_taylor_expand}, we find
\begin{equation}
    \beta
    =
    \frac{\text{num}_\beta}{\text{denom}_\beta}
    \;\propto\;
    \frac{\mathcal{O}(\delta^2)}{\mathcal{O}(\delta)}
    =
    \mathcal{O}(\delta),
\end{equation}
and in the limit of vanishing $\delta$
\begin{equation}
    \lim_{\delta \to 0} \beta = 0.
\end{equation}

An analogous argument applies to $\alpha$. From the analysis above we already know that
$\text{num}_\alpha \propto \mathcal{O}(\delta^2)$, and the denominator of $\alpha$ differs from
$\text{denom}_\beta$ only by a constant factor of $2$, so that
$\text{denom}_\alpha \propto \mathcal{O}(\delta)$ as well. Hence
\begin{equation}
    \alpha
    =
    \frac{\text{num}_\alpha}{\text{denom}_\alpha}
    \;\propto\;
    \frac{\mathcal{O}(\delta^2)}{\mathcal{O}(\delta)}
    =
    \mathcal{O}(\delta),
\end{equation}
and in the limit of vanishing $\delta$ we likewise obtain
\begin{equation}
    \lim_{\delta \to 0} \alpha = 0.
\end{equation}
Therefore, in the limit $\delta \to 0$ both correction coefficients vanish, $\alpha,\beta \to 0$, and the proposed scheme reduces exactly to APEC.

\section{Results}
\label{sec:results}
We now present multiple varying interface problems with a uniform pressure initialization and various equation of states to show the pressure equilibrium properties of the proposed schemes.

\subsection{1D smooth interfaces advection problem - calorically perfect gas}

We begin with the simulation of a calorically perfect gas. Such a gas is governed by the equation of state

\begin{equation}
    \rho e = \frac{P}{\bar{\gamma} - 1},
\end{equation}
where $\bar{\gamma}$ is the specific heat ratio of the mixture which can be written as 

\begin{equation}
    \frac{1}{\bar{\gamma} - 1} = \bar{M} \sum_{i=0}^N \frac{1}{\gamma_i-1} \frac{Y_i}{M_i}
\label{eq:gamma_bar}
\end{equation}

where $N$ is the number of species, $\gamma_i$ and $M_i$ are the ratio of specific heats and the molecular weight of species $i$ respectively.

We run a smooth two–fluid interface test where fluid 1 is a diatomic (air-like) species with $\gamma=1.4$ and $M=28,\mathrm{g/mol}$, and fluid 2 is a monatomic (helium-like) species with $\gamma=1.66$ and $M=4,\mathrm{g/mol}$.

The mass fractions start from a sinusoid, while velocity and pressure are set to uniform:
\begin{equation}
\begin{pmatrix}
(\rho Y_1)_0\\
(\rho Y_2)_0\\
u_0\\
p_0
\end{pmatrix}
=
\begin{pmatrix}
\displaystyle \frac{w_1}{2}\left(1-\sin\bigl(k\,x\right)\Bigr)+0.1\\[6pt]
\displaystyle \frac{w_2}{2}\left(1+\sin\!\left(k\,x\right)\right)+0.1\\[4pt]
1.0\\
0.9
\end{pmatrix},
\label{eq:ideal_gas_ICs}
\end{equation}
with $(w_1,w_2)=(0.6,\,0.2)$ and $k=4\pi$. We use a 4th order Runge-Kutta time integration scheme with a CFL $0.6$ and a periodic domain $x\in[0,1]$ with $501$ grid points. The setup starts in exact pressure/velocity equilibrium, so any drift we see later is purely numerical.

\begin{figure}[h]
    \centering
    \includegraphics[width=0.8\linewidth]{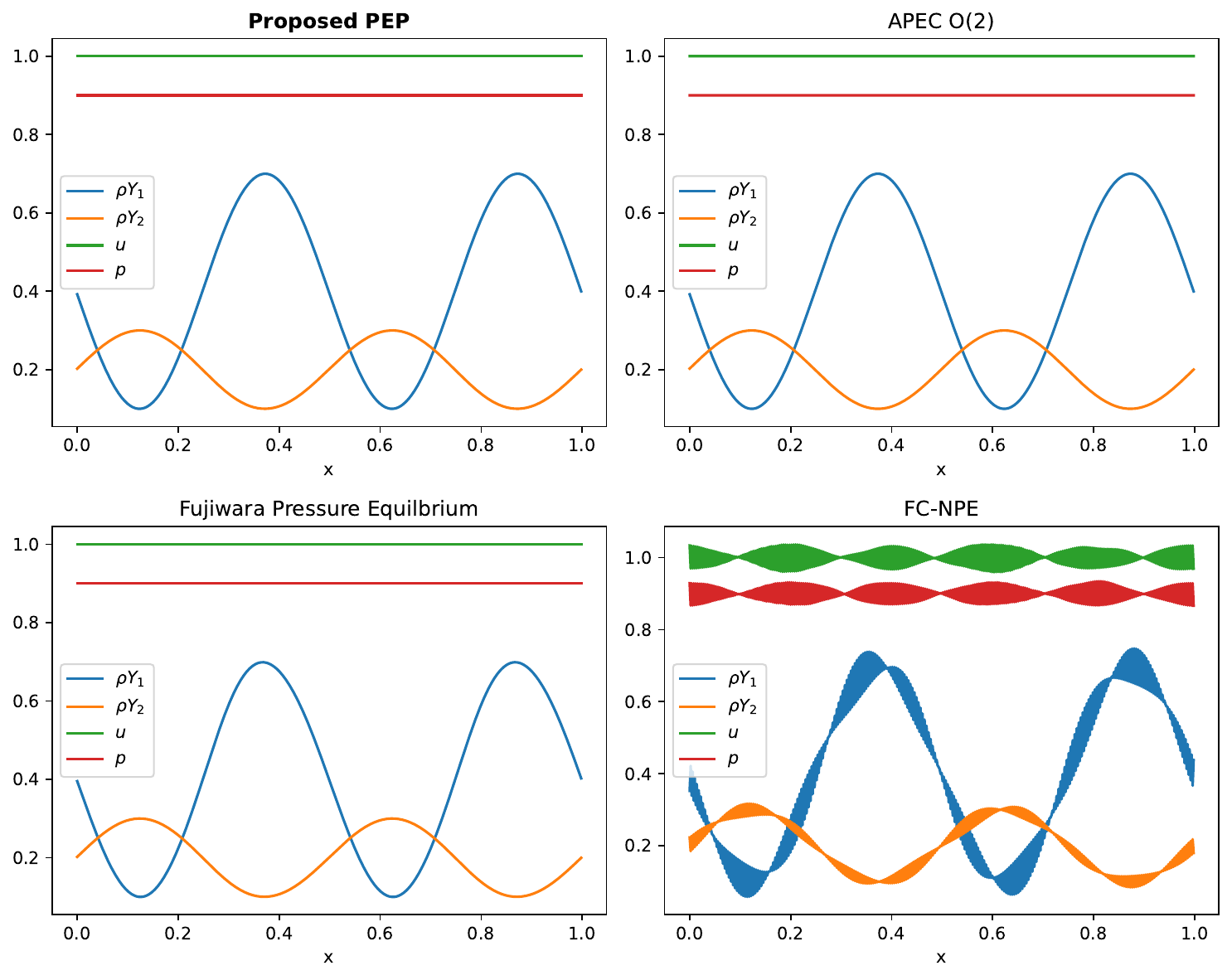}
    \caption{
    Second-order numerical solutions at \(t=20\) for the ideal-gas test case defined in
    Equation~\ref{eq:ideal_gas_ICs}. The four panels show:
    (\textit{top-left}) the proposed PEP scheme;
    (\textit{top-right}) the APEC scheme of~\cite{TERASHIMA2025113701};
    (\textit{bottom-left}) the PEP method of~\cite{fujiwara_fully_2023}; and
    (\textit{bottom-right}) the second-order KEEP scheme (FC-NPE).
    In each panel, blue, orange, green, and red denote
    \(\rho Y_1\), \(\rho Y_2\), \(u\), and \(p\), respectively.
    }
    \label{fig:ideal_profiles}
\end{figure}

Figure \ref{fig:ideal_profiles} shows four second-order solutions at $t=20$ of the ideal-gas test case described in Equation \ref{eq:ideal_gas_ICs}. The top-left is the proposed PEP scheme in Equations \ref{eq:proposed_pep_beg}-\ref{eq:proposed_pep_end}, top-right is APEC as proposed in \cite{TERASHIMA2025113701}, bottom-left is the Fujiwara \cite{fujiwara_fully_2023} PEP scheme, and the bottom-right is the $\mathcal{O}(2)$ KEEP scheme which we will refer to as FC-NPE, fully conservative non pressure equilbrium. The first three schemes show no sign of spurious pressure oscillations while the FC-NPE scheme is being dominated by pressure error at this time.

\begin{figure}[h]
    \centering
    \includegraphics[width=0.45\linewidth]{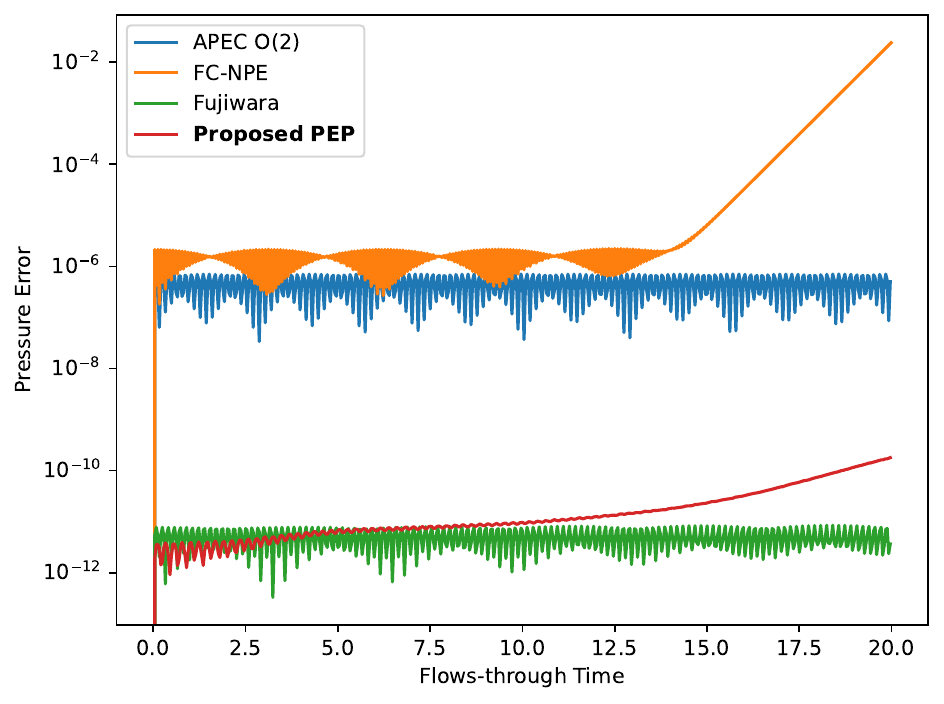}
    \includegraphics[width=0.45\linewidth]{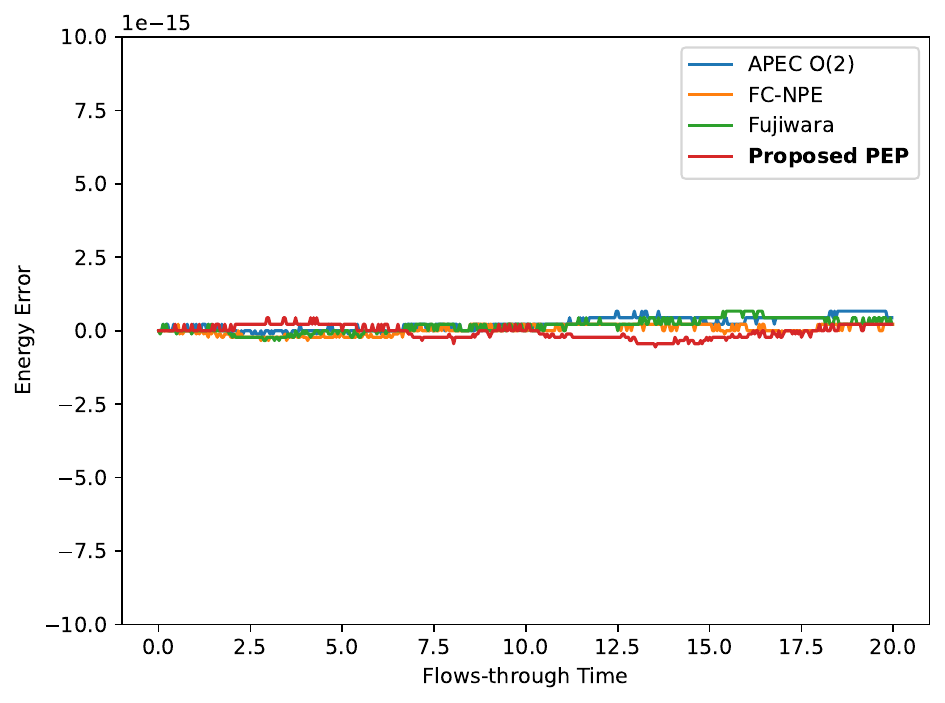}
    \caption{
    \(L_2\) pressure error (left) and total–energy error (right) for the
    ideal-gas test case. Curves correspond to the APEC \(\mathcal{O}(2)\) scheme (blue),
    the FC-NPE scheme (orange), the Fujiwara PEP scheme (green), and the proposed PEP scheme (red).
    }
    \label{fig:ideal_pressure_energy_error}
\end{figure}

To track pressure error quantitatively, we use a relative $L_2$ pressure error over the $N$ interior cells:
\begin{equation}
\varepsilon_p(t) = \sqrt{\frac{1}{N}\sum_{i=1}^{N}\left(\frac{p_i(t)}{p_0}-1\right)^2}
\label{eq:pressure_error}
\end{equation}
In Figure~\ref{fig:ideal_pressure_energy_error}, the pressure error of FC-NPE (orange) grows quickly. APEC $\mathcal{O}(2)$ stays around $10^{-6}$. The Fujiwara scheme is spatially exact and holds near $\approx 10^{-12}$. Our PEP scheme also remains near $10^{-12}$ with a small increase at later times due to an accumilation of errors introduced by the pseudoinverse switch that toggles between APEC and PE behavior.

We also monitor total energy using the same relative $L_2$ formulation:
\begin{equation}
\varepsilon_E(t) = \sqrt{\frac{1}{N}\sum_{i=1}^{N}\left(\frac{E_i(t)}{E_0}-1\right)^2}
\label{eq:energy_error}
\end{equation}
Figure~\ref{fig:ideal_pressure_energy_error} shows energy is conserved for all methods.

\begin{figure}
    \centering
    \includegraphics[width=\linewidth]{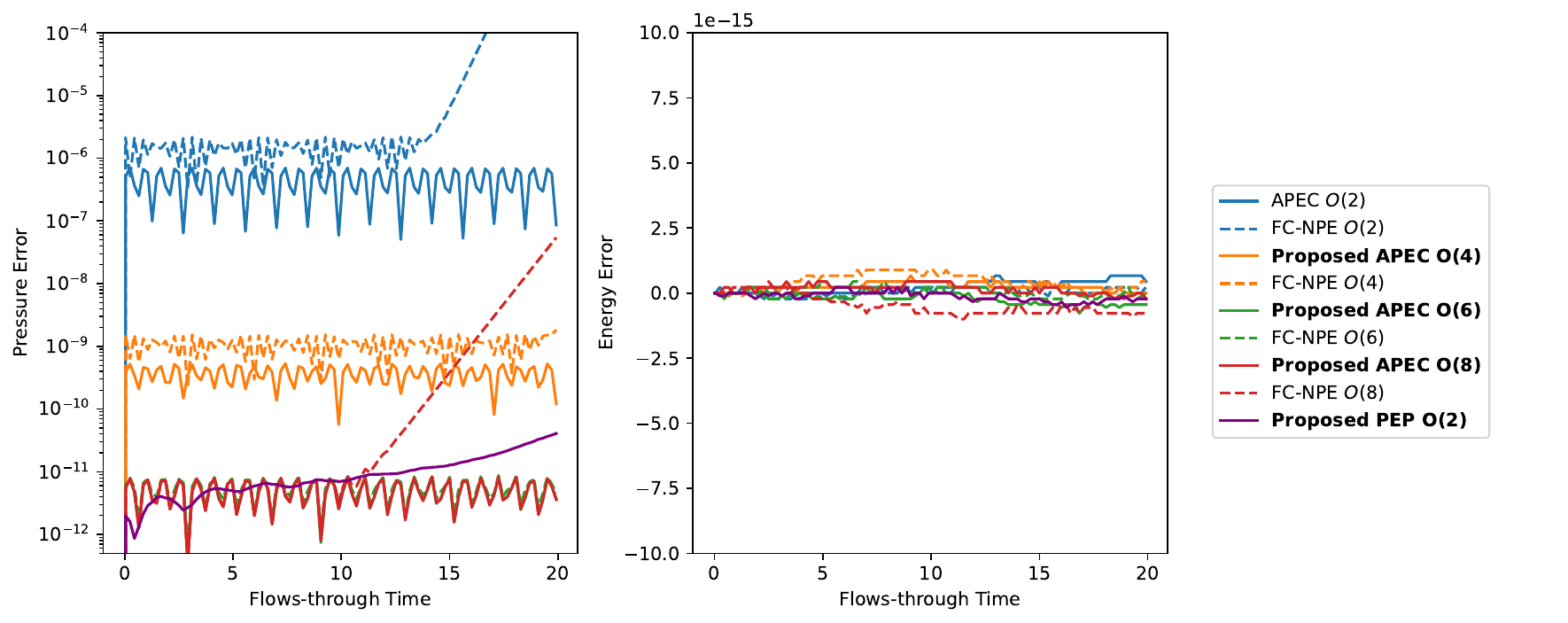}
    \caption{
    \(L_2\) pressure error (left) and total–energy error (right) for
    the higher–order APEC and NPE schemes applied to the ideal-gas test.
    APEC curves are plotted as solid lines and NPE curves as dashed lines, with
    colors denoting the formal order of accuracy:
    \(\mathcal{O}(2)\) (blue), \(\mathcal{O}(4)\) (orange), \(\mathcal{O}(6)\) (green), and \(\mathcal{O}(8)\) (red).
    The proposed PEP scheme is shown in purple for reference.
    \label{fig:apec_vs_npe_ideal}
    }
\end{figure}

We next compare the higher-order formulations in Figure~\ref{fig:apec_vs_npe_ideal}, which shows relative \(L_2\) pressure error (left) and total-energy error (right) for APEC and KEEP (NPE) at second, fourth, sixth, and eighth order. Higher accuracy comes with wider stencils, and for APEC each increase in formal order yields a clear reduction in pressure error. Because APEC is approximate, it does not enforce pressure equilibrium exactly; instead, the imbalance decreases systematically with order, becoming very small by about \(\mathcal{O}(6)\), close to the exact PEP baseline. In contrast, the corresponding NPE variants exhibit growing pressure error at higher orders, highlighting that pressure-equilibrium preservation remains important even for wide-stencil, high-order schemes. On the energy side, all methods are fully conservative, so the relative errors reported in the right panel fluctuate around zero at the level of numerical precision. Notably, the APEC \(\mathcal{O}(2)\) scheme is already very stable, and the higher-order APEC variants inherit this robustness while further reducing the pressure error, with the \(\mathcal{O}(6)\) and \(\mathcal{O}(8)\) forms essentially at numerical precision and remaining accurate for longer times than the second-order PEP scheme, albeit at the cost of a larger stencil and higher formal order. We will no longer show energy-conservation plots for subsequent test cases, as they provide no additional insight once full conservation is verified.

\subsection{1D smooth interfaces advection problem - Stiffened Gas}

We now consider a multi-species stiffened-gas \cite{harlow2004fluid, saurel_abgrall_1999_sisc} where the equation of state can be written as
\begin{equation}    
  \rho e \;=\; \frac{p}{\bar{\gamma}-1} \;+\; \bar{A} \;+\; \rho\,\bar{Q},
\label{eq:stiff_eos}
\end{equation}
where $\bar{\gamma}$ is defined in Equation~\ref{eq:gamma_bar}, $\bar{A}$ is the mixture stiffening-pressure energy offset (pseudo-pressure contribution), defined as
\begin{equation}
  \bar{A}
  \;=\;
  \bar{M}\,\sum_{i=0}^N \frac{\gamma_i\,p_{\infty,i}}{\gamma_i-1}\,\frac{Y_i}{M_i},
\end{equation}
where $p_{\infty,i}$ the stiffening pressure of species $i$. $\bar{Q}$ is the mixture reference specific internal energy (heat of formation) contribution, defined as
\begin{equation}
  \bar{Q}
  \;=\;
  \bar{M}\,\sum_{i=0}^N \frac{q_i}{M_i}\,Y_i.
\end{equation}
where $q_i$ an energy offset of species $i$. 

We define the two fluids as: fluid~1 (water-like) with $\gamma_1=3$, $p_{\infty,1}=0.1$, $q_1=-0.1$, $M_1=18\,\mathrm{g/mol}$; and fluid~2 (air-like) with $\gamma_2=1.4$, $p_{\infty,2}=0$, $q_2=0$, $M_2=29\,\mathrm{g/mol}$. We consider the same initial-value problem as in Eq.~\ref{eq:ideal_gas_ICs}, taking $(w_1,w_2)=(0.6,\,0.2)$ and $k=4\pi$. Time integration uses classical RK4 with CFL $0.6$ on a periodic domain $x\in[0,1]$ with $501$ grid points.

Note that, to the authors' knowledge, there is no fully conservative PEP scheme that accommodates $q_i\neq 0$ other than the one proposed here. The formulation of \cite{fujiwara_fully_2023} covers both ideal-gas and stiffened-gas EOS but is not compatible when the $q$ parameter is nonzero. We therefore deliberately choose one fluid with $q\neq 0$ to expose the consequences of this incompatibility and to assess the behavior of the proposed method, APEC, and a non-pressure-equilbrium (NPE/KEEP) baseline under identical conditions.

\begin{figure}[h]
    \centering
    \includegraphics[width=0.8\linewidth]{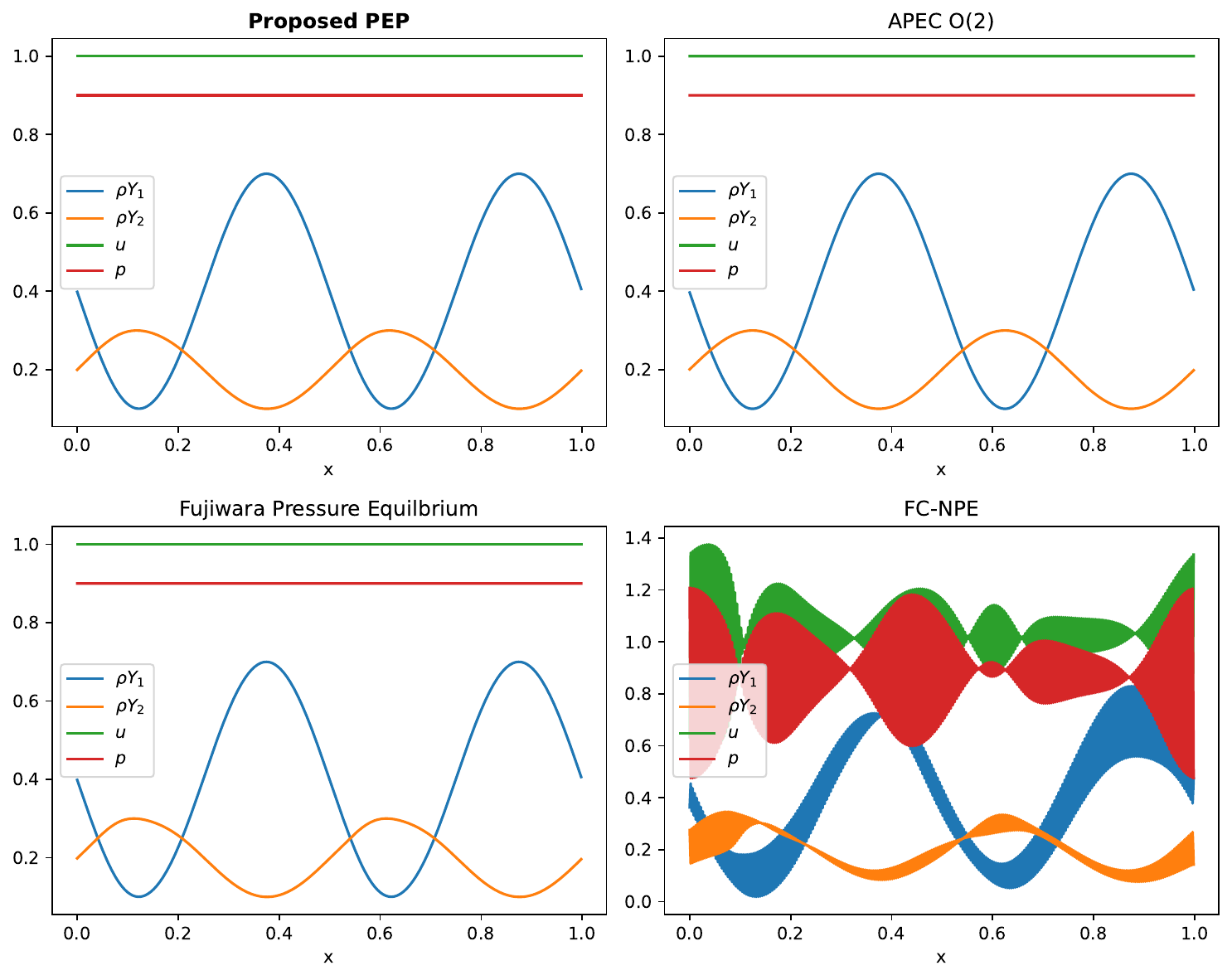}
    \caption{
    Second-order numerical solutions at \(t=7\) for the stiffened-gas test case defined in
    Equation~\ref{eq:ideal_gas_ICs}. The four panels show:
    (\textit{top-left}) the proposed PEP scheme;
    (\textit{top-right}) the APEC scheme of~\cite{TERASHIMA2025113701};
    (\textit{bottom-left}) the PEP method of~\cite{fujiwara_fully_2023}; and
    (\textit{bottom-right}) the second-order KEEP scheme (FC-NPE).
    In each panel, blue, orange, green, and red denote
    \(\rho Y_1\), \(\rho Y_2\), \(u\), and \(p\), respectively.
    }
    \label{fig:stiffened_profile_comparison}
\end{figure}

Figure~\ref{fig:stiffened_profile_comparison} shows the same four second-order methods as before. In the bottom-right panel, the NPE solution exhibits clear pressure oscillations. The other three methods look clean at this scale, with no visible pressure artifacts.

\begin{figure}
    \centering
    \includegraphics[width=0.7\linewidth]{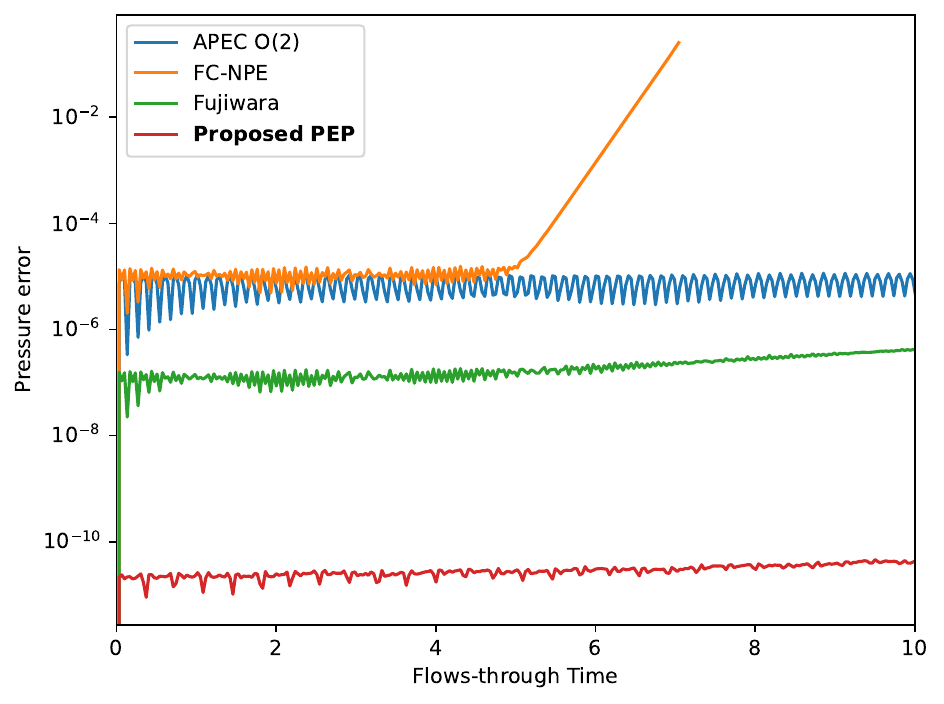}
    \caption{
    \(L_2\) pressure error for the
    stiffened-gas test case. Curves correspond to the APEC \(\mathcal{O}(2)\) scheme (blue),
    the FC-NPE scheme (orange), the Fujiwara PEP scheme (green), and the proposed PEP scheme (red).
    }
    \label{fig:stiffened_pressure_error}
\end{figure}

Figure~\ref{fig:stiffened_pressure_error} zooms in on the pressure equilibrium error. The Fujiwara scheme is no longer preserving pressure equilibrium at numerical precision because it is incompatible with $q\neq 0$. As in the ideal-gas case, APEC reduces error relative to NPE, while NPE quickly diverges. Our proposed PEP method demonstrates the best long-term pressure equilibrium, with a slight increase in pressure error due to the Moore–Penrose pseudoinverse is which can introduce some pressure error as well as small oscillations, consistent with the behavior observed in the ideal-gas case.

\begin{figure}
    \centering
    \includegraphics[width=0.8\linewidth]{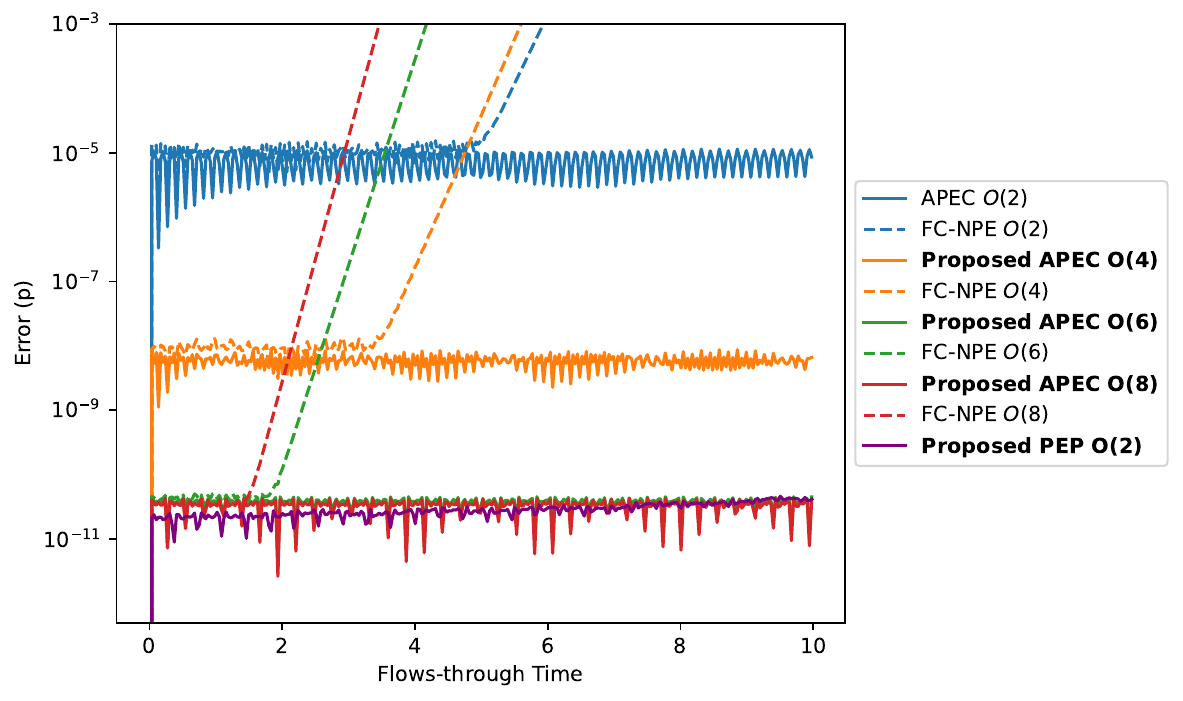}
    \caption{
    \(L_2\) pressure error for
    the higher–order APEC and NPE schemes applied to the stiffened-gas test.
    APEC curves are plotted as solid lines and NPE curves as dashed lines, with
    colors denoting the formal order of accuracy:
    \(\mathcal{O}(2)\) (blue), \(\mathcal{O}(4)\) (orange), \(\mathcal{O}(6)\) (green), and \(\mathcal{O}(8)\) (red).
    The proposed PEP scheme is shown in purple for reference.}
    \label{fig:apec_vs_npe_stiff}
\end{figure}

Figure~\ref{fig:apec_vs_npe_stiff} compares the higher–order APEC and NPE (KEEP) formulations for the stiffened-gas system. As with the ideal-gas case, the NPE schemes begin with modest pressure error at low order, but this error eventually grows and dominates the solution, leading to rapid degradation at later times. In contrast, all APEC variants remain stable, and each increase in formal order yields a corresponding reduction in pressure imbalance. The $\mathcal{O}(6)$ and $\mathcal{O}(8)$ APEC schemes achieve errors approaching numerical precision and maintain this accuracy over long integration times, outlasting the proposed second-order PEP method, whose oscillations arise from the pseudoinverse-based switching mechanism.

\subsection{1D smooth interfaces advection problem — van der Waals}

For the last test problem we adopt a van der Waals mixture EOS with classical mixing rules. The EOS is
\begin{equation}
\left(p + \frac{\bar a}{v^2}\right)\left(v - \bar b\right) = R\,T,
\end{equation}
where $v$ is the specific volume and $R$ is the universal gas constant. The mole fraction of species $i$ is
\begin{equation}
X_i = \frac{Y_i/M_i}{\sum_{j=1}^{N} Y_j/M_j},
\end{equation}
with $N=2$ in this study. The mixture parameters are
\begin{equation}
\bar a = X_1^2\,a_{11} + 2 X_1 X_2\,a_{12} + X_2^2\,a_{22},
\end{equation}
\begin{equation}
\bar b = X_1\,b_1 + X_2\,b_2.
\end{equation}
The pure–component constants are taken from critical data,
\begin{equation}
a_{ii} = \frac{27\left(R T_{c,i}\right)^2}{64\,P_{c,i}},
\end{equation}
\begin{equation}
b_i = \frac{R T_{c,i}}{8\,P_{c,i}},
\end{equation}
and the cross term follows
\begin{equation}
a_{12}=a_{21}=\sqrt{a_{11}a_{22}}.
\end{equation}

We now rewrite the EOS in terms of the compressibility factor $Z \equiv p\,v/(R T)$. Introducing the dimensionless parameters
\begin{equation}
a = \frac{\bar a\,P}{R^2 T^2},
\end{equation}
\begin{equation}
b = \frac{\bar b\,P}{R T},
\end{equation}
the relation becomes a cubic in $Z$,
\begin{equation}
Z^3 - (1+b)\,Z^2 + a\,Z - a\,b = 0.
\end{equation}
At each grid point we solve this cubic for $Z$ and select the physically admissible real root with $Z>b$ (ensuring $v-b>0$). The specific volume then follows from
\begin{equation}
v = \frac{Z\,R\,T}{P}.
\end{equation}

We now consider a two–fluid mixture representative of nitrogen and oxygen, governed by a van der Waals equation of state. The species parameters are
\begin{equation}
\gamma = [1.4,\,1.4],\quad
M = [28,\,32]~\mathrm{g/mol},\quad
T_c = [126.2,\,154.5]~\mathrm{K},\quad
P_c = [3.39,\,5.04]~\mathrm{MPa}.
\end{equation}
On a periodic domain $x\in[0,1]$ with $N=100$ cells, we initialize smooth composition and temperature variations at uniform pressure and velocity. The primitive initial conditions are
\begin{equation}
\begin{pmatrix}
Y_{1,0}\\
Y_{2,0}\\
u_0\\
T_0\\
p_0
\end{pmatrix}
=
\begin{pmatrix}
\displaystyle 0.4\,\sin\!\bigl(2\pi x\bigr) + 0.5\\[6pt]
\displaystyle 1 - Y_{1,0}\\[4pt]
100\\[4pt]
\displaystyle 600 - 300\,\sin\!\bigl(2\pi x\bigr)\\[6pt]
60\times 10^{5}
\end{pmatrix},
\label{eq:vdw_ICs}
\end{equation}

where $p_0$ is in $\mathrm{Pa}$ and $T_0$ in $\mathrm{K}$. At each grid point we compute the mole fractions $X_i$, assemble the mixture van der Waals parameters $\bar a$ and $\bar b$, form $a$ and $b$, solve the cubic for $Z$ with the $Z>b$ selection, recover $v$, and then construct the conservative variables using the van der Waals relations.

In this configuration the flow is again initialized in exact pressure and velocity
equilibrium, but with strong coupled variations in composition and temperature. The
van der Waals EOS makes the thermodynamic response particularly sensitive to these
variations, so this test provides a test case on how each scheme balances
pressure equilibrium against errors in composition and temperature.

\begin{figure}
    \centering
    \includegraphics[width=0.8\linewidth]{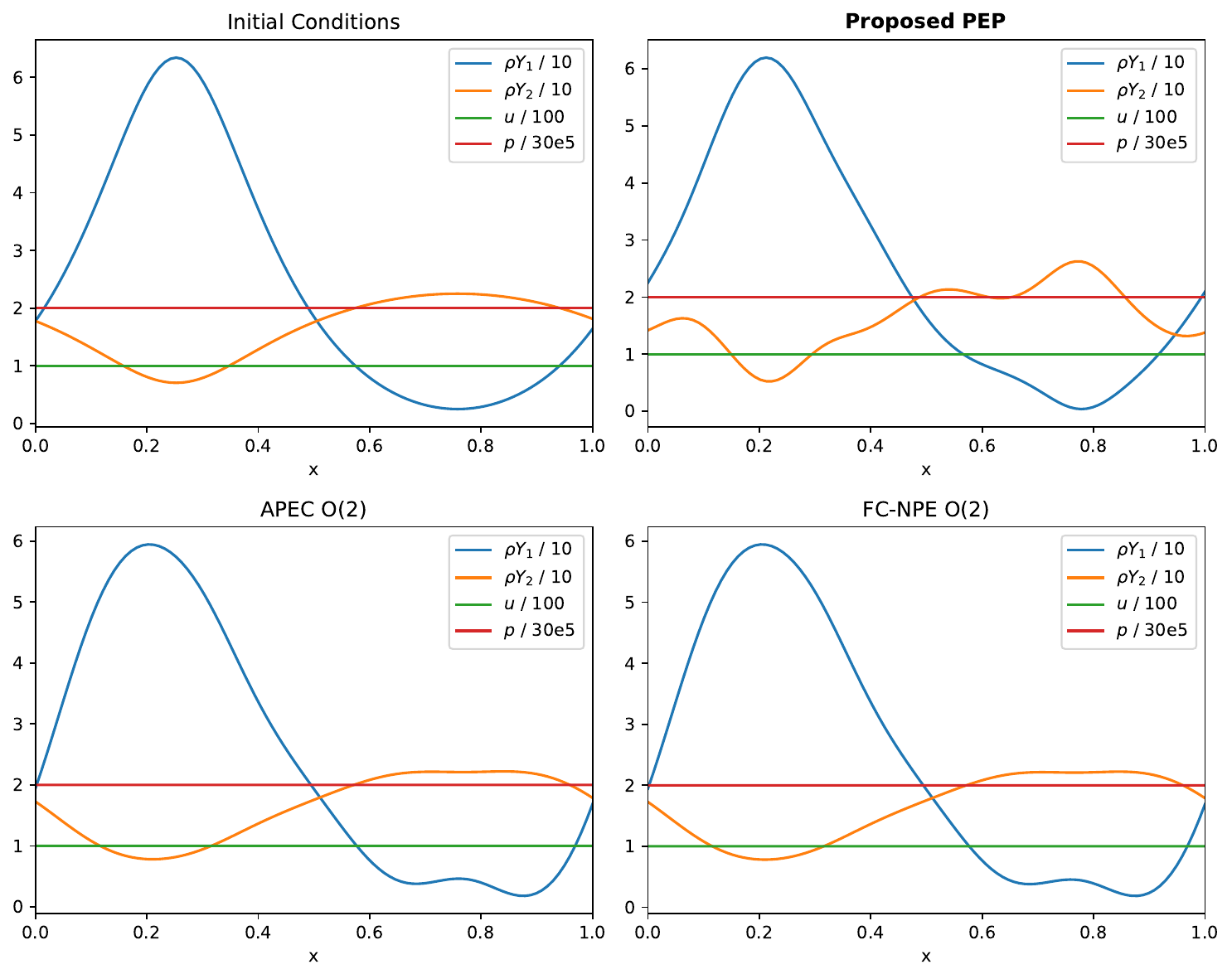}
    \caption{
    Normalized profiles at \(t=0.2\) for the van der Waals mixture test.
    The four panels show:
    (\textit{top-left}) the initial condition;
    (\textit{top-right}) the proposed PEP scheme;
    (\textit{bottom-left}) the APEC \(\mathcal{O}(2)\) scheme; and
    (\textit{bottom-right}) the second-order NPE (KEEP) baseline.
    In each panel, \(\rho Y_1\), \(\rho Y_2\), \(u\), and \(p\) are scaled as indicated
    in the legend to fit on common axes and are colored blue, orange, green, and red respectively. 
    }
    \label{fig:vdw_quad}
\end{figure}

\begin{figure}
    \centering
    \includegraphics[width=0.8\linewidth]{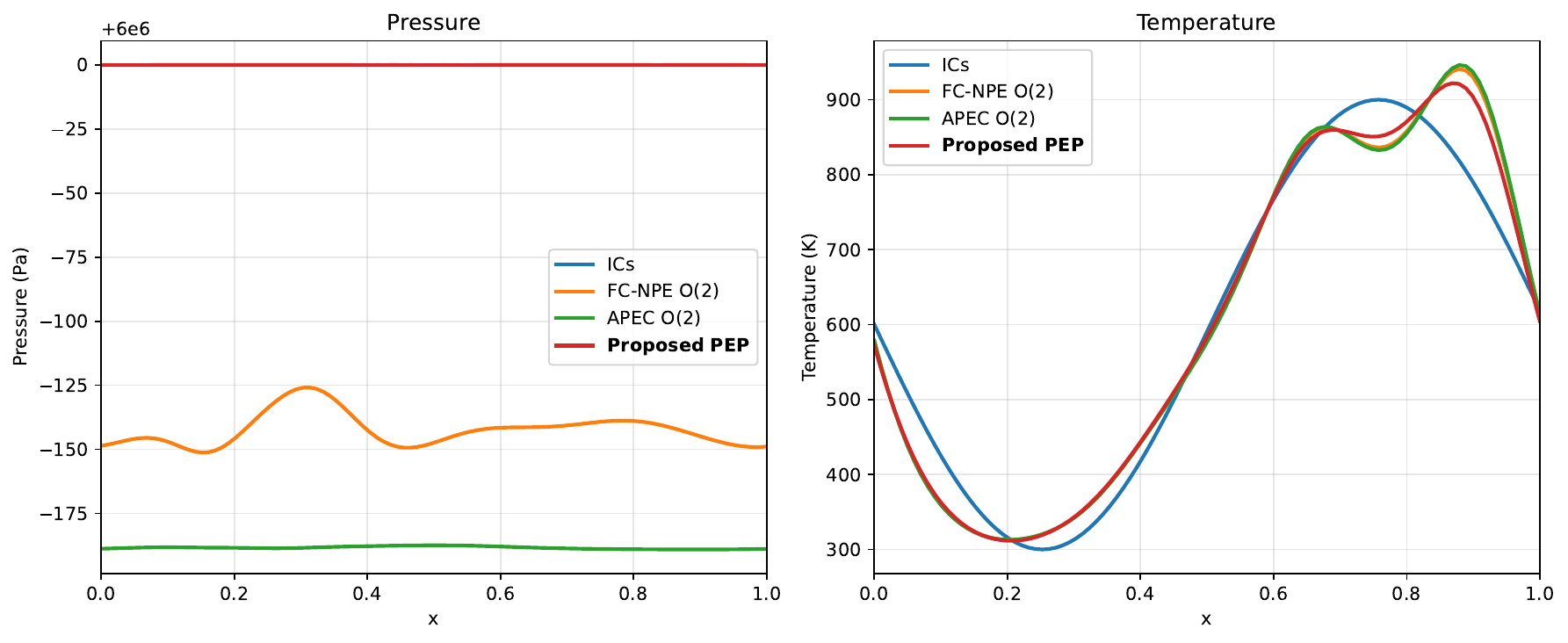}
    \caption{
    Pressure (left) and temperature (right) profiles at \(t=0.2\) for the
    van der Waals test case, comparing the initial condition, the proposed
    PEP scheme, APEC \(\mathcal{O}(2)\), and the second-order NPE (KEEP) baseline.
    Colors: blue – initial condition, orange – NPE, green – APEC \(\mathcal{O}(2)\),
    red – proposed PEP.
    }
    \label{fig:vdw_p_T}
\end{figure}

Figure~\ref{fig:vdw_quad} shows normalized profiles for the second order schemes at \(t=0.2\) (corresponding to
20 flow–through times). The top-left panel shows the
initial condition, while the top-right panel displays the proposed PEP solution, the
bottom-left panel the APEC \(\mathcal{O}(2)\) scheme, and the bottom-right panel the second-order
NPE (KEEP) baseline.  All methods exhibit some distortion of the advected composition
profiles after such long integration times, with noticeable amplitude loss and slight
phase shifts relative to the initial data. A distinctive feature of the proposed PEP
solution is the appearance of new peaks in the \(\rho Y_2\) profile near
\(x \approx 0.8\). This arises because the method enforces pressure equilibrium by
correcting both \(\rho Y_1\), \(\rho Y_2\), and the energy simultaneously, thereby
spreading the thermodynamic correction across multiple fields rather than concentrating
it in a single variable. In contrast, APEC and NPE both show smoother \(\rho Y_2\)
behavior near this location but at the price of a larger pressure error and modifications to the temperature profile as shown in Figure \ref{fig:vdw_p_T}. This shows the pressure (left) and temperature (right) profiles at \(t=0.2 \) s. In the pressure panel, the
proposed PEP solution is visually indistinguishable from the initial condition which indicates the scheme is working as designed to maintain pressure equilbrium. Both APEC \(\mathcal{O}(2)\) and the NPE baseline drift away from the initial uniform pressure: NPE develops a more oscillatory profile but interestingly APEC deviates further away, but with a smoother profile. In the temperature panel, all schemes demonstrates modification to the initial profile due to the need for correction to maintain pressure equilibrium. However,
the proposed PEP scheme again exhibits the smallest departure from the initial
temperature profile (at the cost of the larger deviations in composition). APEC uses the temperature (via the energy convective transport term) to correct pressure error and therefore has the largest deviation in temperature.

\begin{figure}
    \centering
    \includegraphics[width=0.7\linewidth]{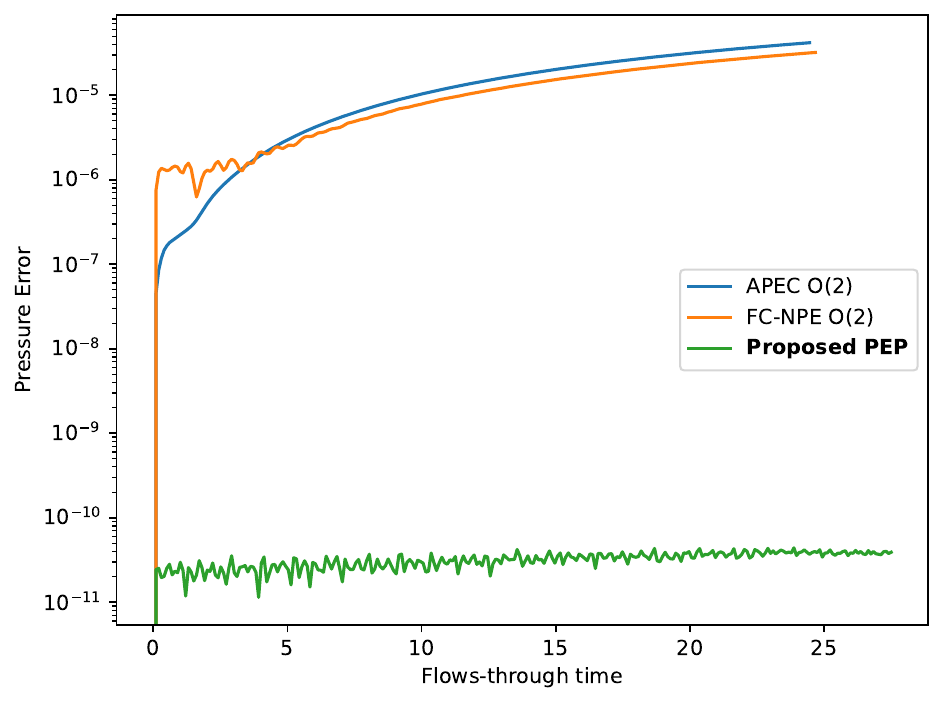}
    \caption{
    Relative \(L_2\) pressure error \(\varepsilon_p(t)\) for the van der Waals test,
    comparing the proposed PEP scheme, APEC \(\mathcal{O}(2)\), and the second-order NPE (KEEP)
    baseline.
    Colors: black – proposed PEP, blue – APEC \(\mathcal{O}(2)\), orange – FC-NPE baseline.
    }
    \label{fig:vdw_pressure_error}
\end{figure}

The time evolution of the pressure–equilibrium error is quantified in
Figure~\ref{fig:vdw_pressure_error}, which plots the relative \(L_2\) pressure error
\(\varepsilon_p(t)\) defined in Equation~\ref{eq:pressure_error} for the proposed PEP
scheme, APEC \(\mathcal{O}(2)\), and the NPE baseline. The proposed PEP method maintains
\(\varepsilon_p\) between \(10^{-11}\) and \(10^{-10}\) throughout the integration,
essentially at the level of numerical precision for this problem. In contrast, both
APEC and NPE are around \(\mathcal{O}(10^{-5})\) with NPE actually having a lower error for later times. This is confirmed by Figure \ref{fig:vdw_p_T} where we see APEC deviate further from initial condition in terms of absolute pressure levels, but maintaining a smoother pressure profile than NPE. 

When further continuing the simulations beyond 20 flow-through times, before any large instability induced by pressure in-equilibrium occurs, all three schemes accumulate dispersion error which eventually cause the simulation to fail when $\rho Y_1$ dips below 0 at $x \approx 0.9$. This trend can be seen coming in Figure \ref{fig:vdw_quad} as all 3 profiles have a small dip in $\rho Y_1$ approaching zero. The proposed PEP scheme shares this downfall with APEC and NPE as central schemes, incurring dispersion error that is not bound preserving. Handling this long–time dispersive behavior with an appropriate regularization, such as a artificial viscosity or filter, is an important direction for future work, but we emphasize that the reason the proposed PEP scheme fails at very long times is not due to pressure error, where a pressure error of $\approx 10^{-10}$ is maintained for the full duration up to the abrupt failure due to negative mass fractions.

\begin{figure}
    \centering
    \includegraphics[width=0.8\linewidth]{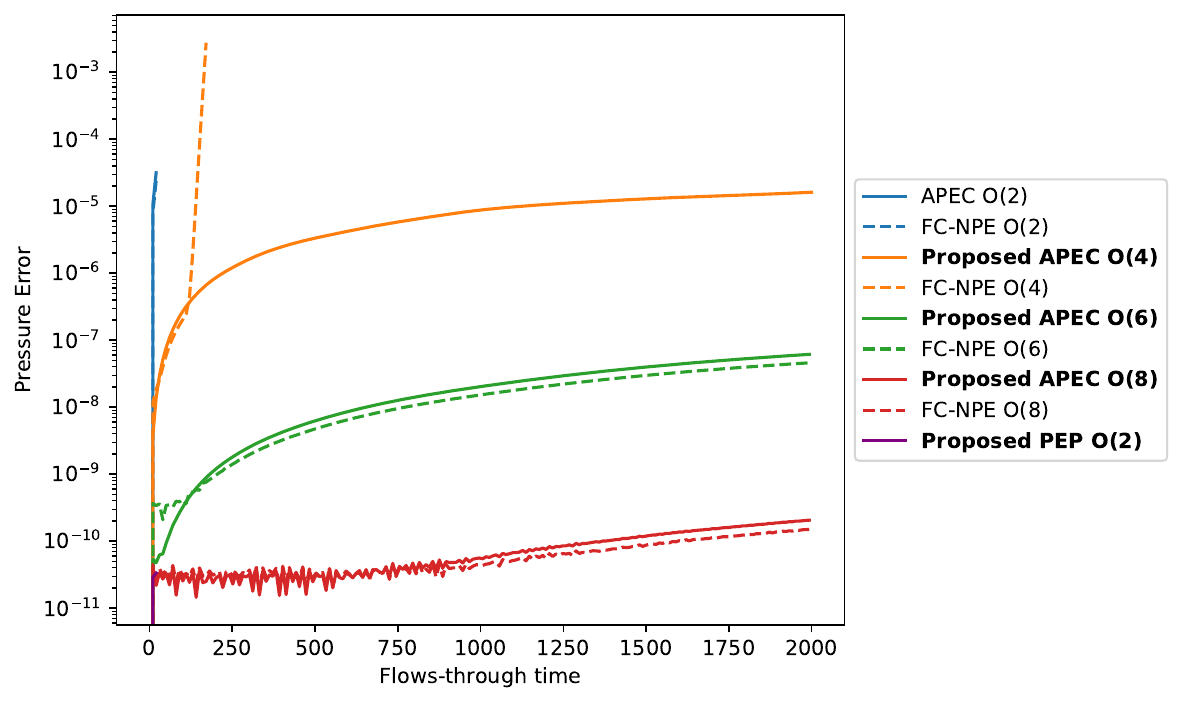}
    \caption{
    Relative \(L_2\) pressure error for higher–order APEC and NPE (KEEP) schemes,
    together with the proposed PEP method, applied to the van der Waals mixture test.
    APEC curves are shown as solid lines and NPE curves as dashed lines.
    Colors indicate formal order and scheme:
    blue – \(\mathcal{O}(2)\), orange – \(\mathcal{O}(4)\), green – \(\mathcal{O}(6)\), red – \(\mathcal{O}(8)\),
    purple – proposed PEP (formally second order).
    }
    \label{fig:vdw_apec_vs_npe}
\end{figure}

Finally, Figure~\ref{fig:vdw_apec_vs_npe} shows the higher–order APEC and NPE (KEEP)
variants, together with the proposed PEP scheme, for the van der Waals test. As in the
ideal-gas and stiffened-gas cases, APEC curves are plotted as solid lines and NPE
curves as dashed lines, with colors denoting the formal order of accuracy
(\(\mathcal{O}(2)\), \(\mathcal{O}(4)\), \(\mathcal{O}(6)\), \(\mathcal{O}(8)\)) as before. We again focus on the relative
\(L_2\) pressure error \(\varepsilon_p(t)\). For this more challenging EOS with
large temperature variations, the wider APEC stencils yield a notable benefit: the
thermodynamic corrections induced by the approximate pressure-equilibrium constraint are
spread over a broader number of points. As a result, the APEC
\(\mathcal{O}(6)\) and \(\mathcal{O}(8)\) schemes remain extremely stable over the time window shown
(\(t\le 20\), i.e.\ roughly 2000 flow-through times), with pressure errors near machine
precision and smooth profiles in both composition and temperature. The NPE \(\mathcal{O}(4)\)
scheme, by contrast, develops significantly larger pressure error growing exponentially after roughly $200$ flows through time.

\section{Conclusion}
This work introduced two complementary strategies for suppressing spurious pressure oscillations in multi-component flows with general equations of state: a high-order extension of the approximate pressure-equilibrium-preserving (APEC) correction, and a fully conservative scheme that enforces the discrete PEP condition exactly. Although APEC remains approximate and cannot preserve pressure equilibrium exactly, its high-order extension substantially improves accuracy, and in practice its pressure errors often approach \emph{exactly} pressure-equilibrium preserving schemes, while retaining excellent robustness. The proposed fully conservative PEP scheme, by contrast, satisfies the PEP criterion exactly for any arbitrary equation of state and preserves all conserved quantities, but its long-time behavior is more sensitive to numerical noise introduced by the pseudo-inverse, leading to a characteristic pressure-error growth distinct from that of previous methods.

These findings highlight several promising directions for future work. Extension of a targeted pressure-equilibrium artificial viscosity \cite{terashima_consistent_2013} using the PEP treatment presented in this paper may help regularize the conservative update and suppress long-time noise amplification introduced by the proposed PEP scheme but could also help with the dispersion error. Likewise, hybrid schemes that blend high-order APEC with the proposed PEP scheme could improve stability without sacrificing accuracy. One could also pursue a genuine high-order extension of the proposed fully-conservative PEP method, in which the exact PEP criterion is enforced across all sub-stencil intersections, mirroring the construction of the high-order APEC scheme.

Finally, the broader impact of this framework lies in its generality. Both the high-order APEC correction and the proposed conservative PEP update act as modular correction steps and can be incorporated into any numerical flux function for any EOS. While this work employed KEEP as a specific starting point, the methodology is not restricted to it. Extensions to solvers relying on traditional Riemann fluxes are straightforward and represent a natural next step. 

In summary, high-order APEC provides a practical and effective tool for mitigating spurious pressure oscillations for arbitrary EOS and multi-species systems, while the fully conservative PEP formulation establishes a flexible and extensible foundation for future development of \emph{exactly} pressure-equilibrium preserving and fully conservative multi-component flow solvers.

\section*{Acknowledgments}
The authors would like to thank Dr. Abram Rodgers and Dr. Chase Ashby for helpful conversations and internal review of this work. Support for this work was provided by the NASA Pathways Program and the Exploration Ground Systems (EGS) Program under the Exploration Systems Development Mission Directorate (ESDMD).

\bibliographystyle{unsrt}
\bibliography{references}

\appendix
\section{Sensitivities to Reciprocal Condition Number}
\label{sec:rcond_sensitivity}

In Section~\ref{sec:proposed_scheme} we introduced a global reciprocal condition number
$r_g$ defined by Equation \ref{eq:rcond_global}, obtained by scanning the local face-wise
condition estimates $\atfaceplus{r}{m}$ at the initial condition. For the three EOS
configurations considered in this work, the resulting values are
\begin{align*}
\text{Ideal gas:} \quad & r_g \approx 1.16\times 10^{-5}, \\
\text{Stiffened gas:} \quad & r_g \approx 3.20\times 10^{-6}, \\
\text{van der Waals:} \quad & r_g \approx 2.46\times 10^{-9}.
\end{align*}
These values reflect the different conditioning properties of the PEP system across equations of state for the fluid systems and ICs presented. It is important to note that these values were for the specific ICs and fluid systems and will not be optimal generally for an arbitrary fluid system or IC with the same EOS. Instead, please use the dynamic procedure established in Equation \ref{eq:rcond_global}.

\subsection{Local behavior of \texorpdfstring{$\alpha$}{alpha} and \texorpdfstring{$\beta$}{beta}}
\label{sec:rcond_alpha_beta}

To build intuition for the role of the reciprocal condition number, we first examine
its effect on the \emph{local} PEP correction coefficients $\alpha$ and $\beta$. We
consider the two-species ideal-gas mixture used in the main text and set up a simple
interface experiment. Starting from uniform pressure and velocity, we prescribe a
species jump
\begin{equation}
\rho Y_{1,m+1} = \rho Y_{1,m} + \Delta(\rho Y_i), \qquad
\rho Y_{2,m+1} = \rho Y_{2,m} + \Delta(\rho Y_i),
\end{equation}
and, for each value of $\Delta(\rho Y_i)$, we compare the closed-form analytic solution in Equations
\ref{eq:alpha}--\ref{eq:beta}, the Moore--Penrose pseudoinverse with several fixed reciprocal condition number values, which we denote,
\texttt{rcond}, and the automatic choice based on the $r_g$ construction in
Equation \ref{eq:rcond_global}. In all cases, the second-order APEC scheme corresponds to
$\alpha=\beta=0$.

\begin{figure}[t]
    \centering
    \includegraphics[width=\textwidth]{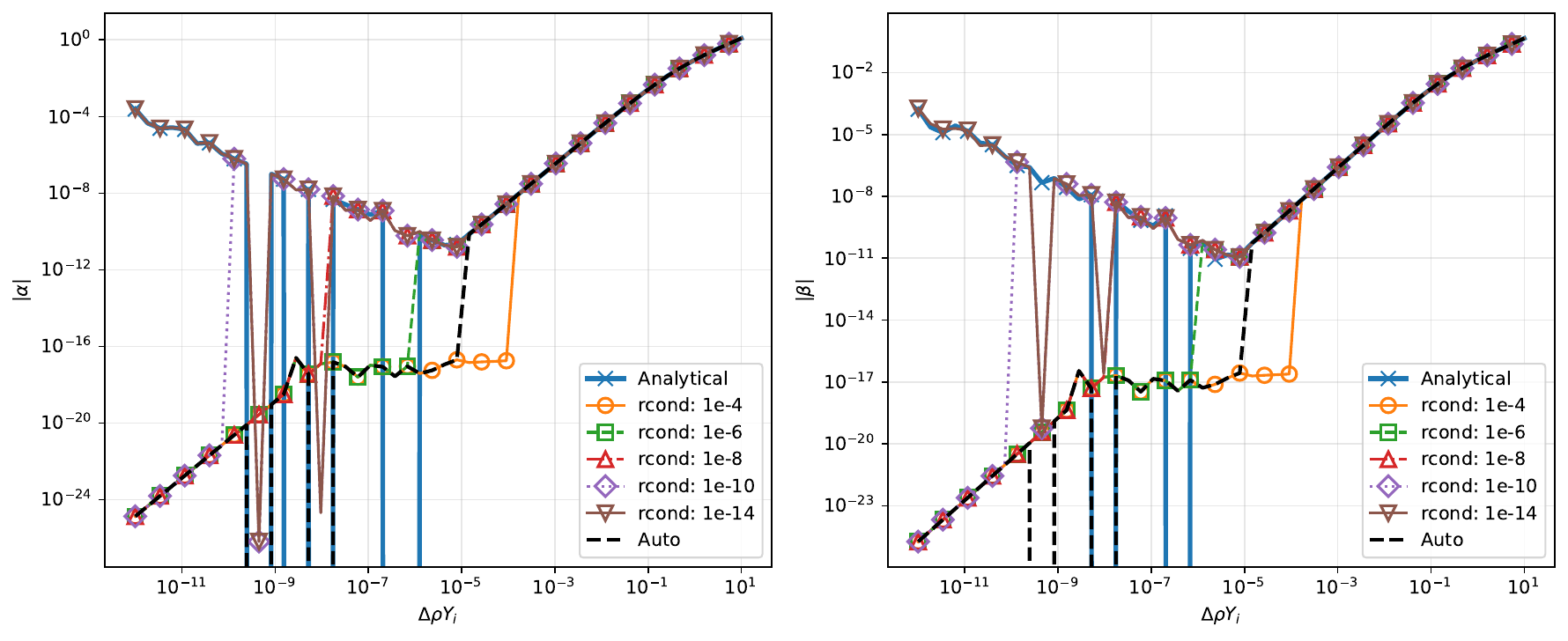}
    \caption{Local sensitivity of the PEP correction coefficients $\alpha$ (left) and
    $\beta$ (right) to the reciprocal condition number in an ideal-gas two-species
    mixture. Each curve shows $\alpha$ or $\beta$ as a function of the species jump
    $\Delta(\rho Y_i)$. The analytic closed-form solution (no regularization) displays
    strong oscillations as $\Delta(\rho Y_i)\to 0$, while the APEC baseline
    corresponds to $\alpha=\beta=0$. The pseudoinverse results with fixed
    \texttt{rcond} values interpolate between these limits, and the automatic
    $r_g$-based choice yields a smooth decay of $\alpha$ and $\beta$ to zero near the
    determinant-limiting regime.}
    \label{fig:ideal_alpha_beta}
\end{figure}

Figure~\ref{fig:ideal_alpha_beta} shows the resulting $\alpha$ (left panel) and
$\beta$ (right panel) as functions of $\Delta(\rho Y_i)$. The
analytic closed-form solution exhibits the strongest oscillations: as
$\Delta(\rho Y_i)\to 0$, the underlying linear system becomes nearly singular, and
roundoff errors are amplified into large, oscillatory variations in $\alpha$ and
$\beta$. Very small tolerances (e.g.\ \texttt{rcond} $=10^{-14}$) closely track this
analytic curve and therefore inherit its oscillations. On the other hand, very large
tolerances (e.g.\ \texttt{rcond} $=10^{-4}$) clamp the correction rapidly toward zero causing the coefficients $\alpha$ and $\beta$ are forced to vanish even for moderately sized
composition jumps, so the scheme behaves like APEC over a wide range of
$\Delta(\rho Y_i)$. We note that we are only concerned with the oscillations above numerical precision. The auto curve exhibits some oscillations between exact zero and numerical zero ($\approx 10^{-16}$ in double precision) in the $\Delta \rho Y_i \approx 10^{-9}$ with no adverse effects. Compared to similar oscillations from the analytical scheme which is oscillating between numerical zero and $\approx 10^{-7}$.

Between these extremes there is clearly an \emph{optimal} range of \texttt{rcond}
values. As $\Delta(\rho Y_i)$ decreases toward zero, the ideal behavior is for the
PEP corrections to decay smoothly to $\alpha=\beta=0$, so that the scheme transitions
gradually from the fully conservative PEP correction to the APEC fallback without
numerical oscillations. In Figure~\ref{fig:ideal_alpha_beta}, intermediate tolerances such
as \texttt{rcond} $=10^{-6}$ achieve this behavior. Larger tolerances (e.g.\ \texttt{rcond} $=10^{-4}$) cut off
this decay too sharply and effectively switch to APEC too early, while smaller
tolerances oscillate in the near-singular regime. For the ideal-gas test, the
best-performing fixed choice \texttt{rcond} $=10^{-6}$ which is the closest to the global, conditioning-based estimate, $r_g \approx 1.16\times 10^{-5}$.

\subsection{Global pressure-error behavior across EOS}
\label{sec:rcond_pressure}

\begin{figure}[t]
    \centering

    \begin{minipage}{0.47\textwidth}
        \centering
        \includegraphics[width=\textwidth]{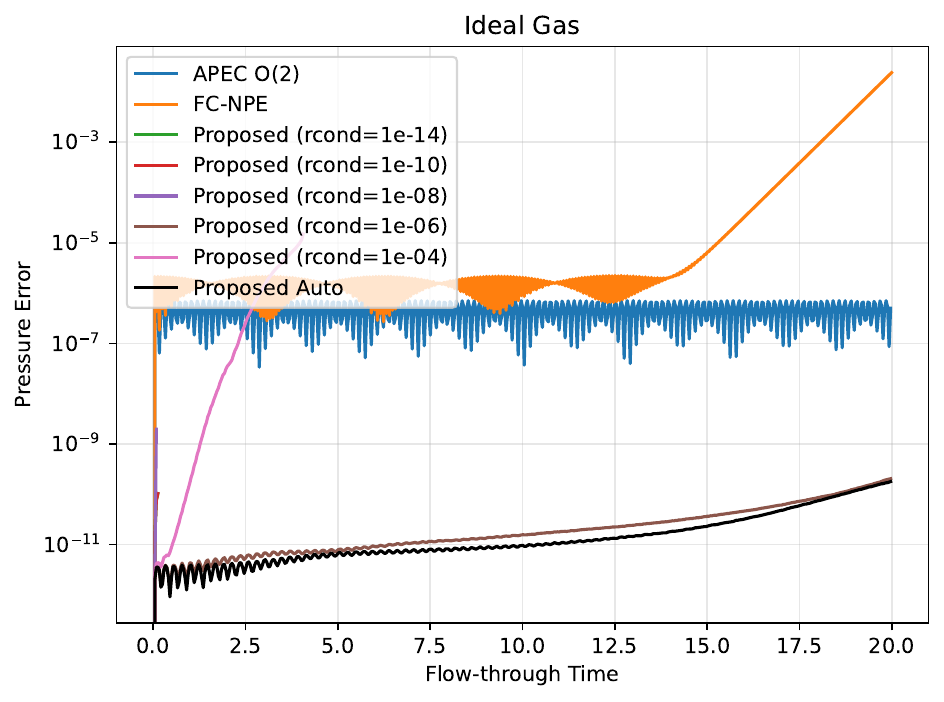}
    \end{minipage}
    \hfill
    \begin{minipage}{0.47\textwidth}
        \centering
        \includegraphics[width=\textwidth]{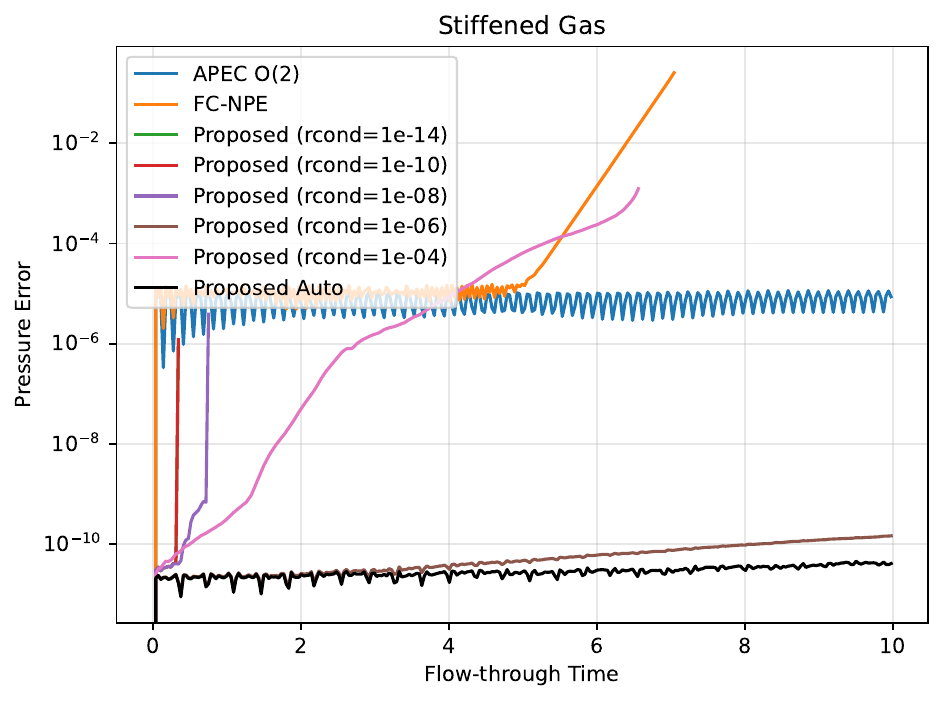}
    \end{minipage}

    \vspace{0.4cm}

    \begin{minipage}{0.47\textwidth}
        \centering
        \includegraphics[width=\textwidth]{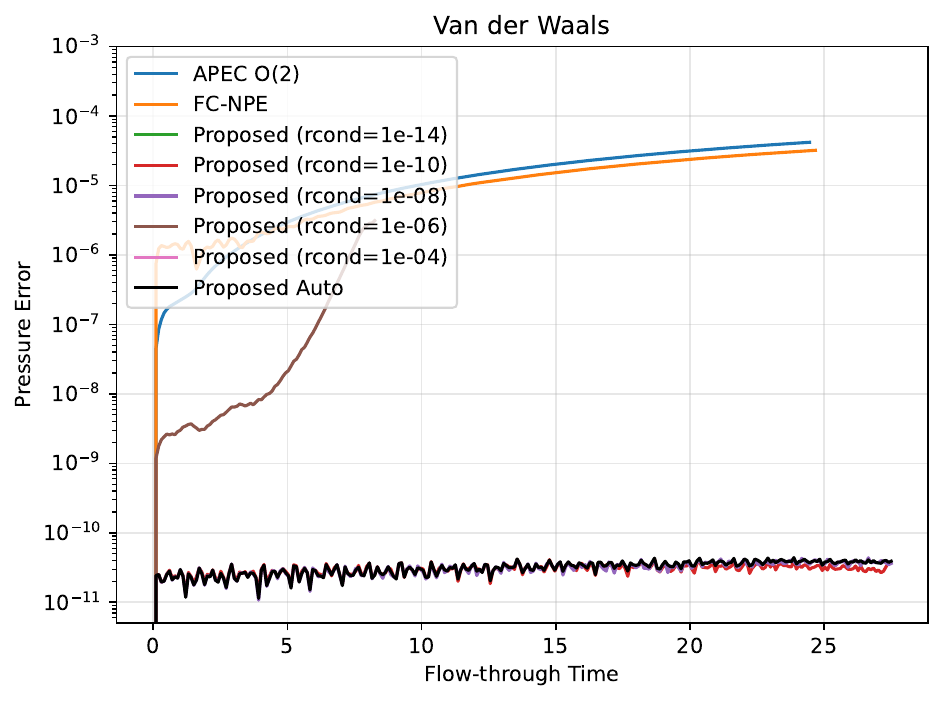}
    \end{minipage}

    \caption{Sensitivity of long-time pressure error to the reciprocal condition number
    used in the Moore--Penrose pseudoinverse for the three EOS configurations
    (ideal-gas, stiffened-gas, van der Waals). Each panel shows the $\ell_1$ pressure
    error versus flow-through time. The second-order APEC scheme and the FC-NPE
    baseline are plotted as references, along with the proposed fully conservative PEP
    scheme for several fixed \texttt{rcond} values and the automatic choice $r_g$ from
    Equation ~\ref{eq:rcond_global}.}
    \label{fig:rcond_pressure_error_all_eos}
\end{figure}

The local picture in \ref{sec:rcond_alpha_beta} manifests at the global level
as a balance between accuracy and robustness in the long-time pressure error. To
quantify this effect, we repeat the ideal-gas, stiffened-gas, and van-der-Waals
numerical experiments using the fully conservative PEP scheme, but vary the reciprocal
condition number used in the Moore--Penrose pseudoinverse. In addition to the
automatic $r_g$ choice from Equation ~\ref{eq:rcond_global}, we consider several fixed
tolerances (here \texttt{rcond} $=10^{-14}$, $10^{-10}$, $10^{-8}$, $10^{-6}$, and
$10^{-4}$) to sample both the under-regularized and over-regularized regimes.
The resulting $\ell_2$ pressure errors as functions of flows-through time are shown in
Figure~\ref{fig:rcond_pressure_error_all_eos}. Each panel includes the second-order
APEC baseline and the fully conservative NPE (FC-NPE) reference, together with the
proposed PEP scheme for the various \texttt{rcond} values and the automatic $r_g$ choice as presented in the main body.

The trends mirror the local behavior of $\alpha$ and $\beta$. When \texttt{rcond} is
too small, the pseudoinverse admits nearly all singular values, including those
associated with ill-conditioned faces. The resulting oscillatory corrections in
$\alpha$ and $\beta$ produce high-frequency noise in the fluxes and a rapid growth of
the global pressure error. When \texttt{rcond} is too large, the solver discards many
small singular values, effectively switching to the APEC flux over a large portion of
the domain; the solution remains robust, but the pressure error grows more quickly and
approaches the APEC baseline.

In between, there is an intermediate range in which the PEP correction is active
where it is most needed (near strong composition jumps) and smoothly deactivated in
the ill-conditioned regime, yielding both low pressure error and good long-time
stability. Across all three EOS, the automatically selected $r_g$ lies in this
intermediate regime and closely tracks the best fixed-\texttt{rcond} choices,
consistent with the local behavior in Figure~\ref{fig:ideal_alpha_beta}. This suggests
that the conditioning-based construction of $r_g$ provides a simple, problem-aware
default that does not require manual tuning, while still delivering a smooth and
physically sensible transition between the fully conservative PEP update and its APEC
fallback.

\section{Extension of the Proposed PEP Scheme to Single Species}

Although the proposed PEP formulation is developed for multi-component flows with
general equations of state, we present a simplified variant for single-species flow. In this case, there is only one conserved species density,
which we denote by $\rho Y = \rho$. The pressure-equilibrium criterion shown in Equation 
\ref{eq:pep_property_final} reduces to corrections on $\rho$ and $\rho e$ only. We can therefore define coefficient correction terms $\alpha'$ and $\beta'$ as

\begin{align}
    \atfaceplus{\rho}{m}^{\mathrm{PEP}}
    &= \frac{\atcell{\rho}{m} + \atcell{\rho}{m+1}}{2} + \alpha', 
    \label{eq:ss_rho} \\[6pt]
    \atfaceplus{\rho e}{m}^{\mathrm{PEP}}
    &= \frac{\atcell{\rho e}{m} + \atcell{\rho e}{m+1}}{2}
    + \beta'
    - \left[
        \frac{\epsilon_{m+1}}{2}(\rho_{m+1} - \rho_{m+\frac12})
        - \frac{\epsilon_m}{2}(\rho_{m+\frac12} - \rho_m)
    \right]
    \label{eq:ss_rhoe}
\end{align}

where we leave the APEC correction in Equation \ref{eq:ss_rhoe} as before so as $\alpha',\beta'$ approach zero the scheme exactly recovers the single-species APEC scheme. The extension to species $>2$ is discussed in Section \ref{sub:nspecies}.

\end{document}